\documentclass[lettersize,journal]{IEEEtran}
\usepackage{setspace}
\usepackage[mathscr]{eucal}
\usepackage{amssymb,amsmath,amsfonts,latexsym}
\usepackage{xcolor,url,overpic}
\usepackage{array}
\usepackage{verbatim}
\usepackage{bm}
\usepackage{bbm}
\usepackage{dsfont}
\usepackage{algorithmic}
\usepackage{algorithm}
\usepackage{verbatim}
\usepackage{textcomp}
\usepackage{mathrsfs}
\usepackage{epstopdf}
\usepackage{booktabs} 
\usepackage{thm-restate}

\newcommand{\openone}{\leavevmode\hbox{\small1\normalsize\kern-.33em1}}

\catcode`~=11 \def\UrlSpecials{\do\~{\kern -.15em\lower .7ex\hbox{~}\kern .04em}} \catcode`~=13 

\allowdisplaybreaks[1]

\newcommand{\nn}{\nonumber}

\newcommand{\calA}{\mathcal{A}}
\newcommand{\calB}{\mathcal{B}}

\newcommand{\calH}{\mathcal{H}}

\newcommand{\calL}{\mathcal{L}}

\newcommand{\calO}{\mathcal{O}}
\newcommand{\calP}{\mathcal{P}}

\newcommand{\calR}{\mathcal{R}}
\newcommand{\calS}{\mathcal{S}}

\newcommand{\calV}{\mathcal{V}}

\newcommand{\calX}{\mathcal{X}}

\newcommand{\calZ}{\mathcal{Z}}

\newcommand{\rmH}{\mathrm{H}}

\newcommand{\bbE}{\mathbb{E}}

\newcommand{\bbR}{\mathbb{R}}

\DeclareMathAlphabet{\mathbsf}{OT1}{cmss}{bx}{n}
\DeclareMathAlphabet{\mathssf}{OT1}{cmss}{m}{sl}

\newcommand{\rvD}{\mathsf{D}}

\DeclareSymbolFont{bsfletters}{OT1}{cmss}{bx}{n}  
\DeclareSymbolFont{ssfletters}{OT1}{cmss}{m}{n}
\DeclareMathSymbol{\bsfGamma}{0}{bsfletters}{'000}
\DeclareMathSymbol{\ssfGamma}{0}{ssfletters}{'000}
\DeclareMathSymbol{\bsfDelta}{0}{bsfletters}{'001}
\DeclareMathSymbol{\ssfDelta}{0}{ssfletters}{'001}
\DeclareMathSymbol{\bsfTheta}{0}{bsfletters}{'002}
\DeclareMathSymbol{\ssfTheta}{0}{ssfletters}{'002}
\DeclareMathSymbol{\bsfLambda}{0}{bsfletters}{'003}
\DeclareMathSymbol{\ssfLambda}{0}{ssfletters}{'003}
\DeclareMathSymbol{\bsfXi}{0}{bsfletters}{'004}
\DeclareMathSymbol{\ssfXi}{0}{ssfletters}{'004}
\DeclareMathSymbol{\bsfPi}{0}{bsfletters}{'005}
\DeclareMathSymbol{\ssfPi}{0}{ssfletters}{'005}
\DeclareMathSymbol{\bsfSigma}{0}{bsfletters}{'006}
\DeclareMathSymbol{\ssfSigma}{0}{ssfletters}{'006}
\DeclareMathSymbol{\bsfUpsilon}{0}{bsfletters}{'007}
\DeclareMathSymbol{\ssfUpsilon}{0}{ssfletters}{'007}
\DeclareMathSymbol{\bsfPhi}{0}{bsfletters}{'010}
\DeclareMathSymbol{\ssfPhi}{0}{ssfletters}{'010}
\DeclareMathSymbol{\bsfPsi}{0}{bsfletters}{'011}
\DeclareMathSymbol{\ssfPsi}{0}{ssfletters}{'011}
\DeclareMathSymbol{\bsfOmega}{0}{bsfletters}{'012}
\DeclareMathSymbol{\ssfOmega}{0}{ssfletters}{'012}

\DeclareMathOperator*{\argmin}{arg\,min}

\newcommand{\Unif}{\mathrm{Unif}}

\usepackage{thmtools, thm-restate}
\newtheorem{theorem}{Theorem} 
\newtheorem{lemma}[theorem]{Lemma}

\newtheorem{definition}{Definition} 
\newtheorem{example}{Example}

\newtheorem{assumption}{Assumption}

\usepackage{mathtools}
\usepackage{amsmath,amsfonts}
\usepackage{algorithmic}
\usepackage{algorithm}
\usepackage{array}
\usepackage[caption=false,font=normalsize,labelfont=sf,textfont=sf]{subfig}
\usepackage[font=normalsize]{caption}
\usepackage{textcomp}
\usepackage{stfloats}
\usepackage{url}
\usepackage{verbatim}
\usepackage{graphicx}
\usepackage{cite}
\usepackage{wrapfig}
\usepackage{tikz}
\usepackage{pgf}
\usetikzlibrary{babel,automata,shapes}
\usepackage{fontawesome5}
\usetikzlibrary{shadows}
\usetikzlibrary{shadows.blur}
\usetikzlibrary{positioning}
\usepackage[table,x11names]{xcolor}
\usepackage{multirow}
\usepackage{makecell}
\usepackage{booktabs}
\usepackage{enumitem}
\mathtoolsset{showonlyrefs}

\newcommand{\cP}{\mathcal{P}}

\newcommand{\cV}{\mathcal{V}}

\newcommand{\cX}{\mathcal{X}}

\newcommand{\sD}{\mathsf{D}}

\newcommand{\TV}{\mathsf{TV}}

\mathchardef\mhyphen="2D

\newcommand{\pt}{\mathrm{pt}}
\newcommand{\fix}{\mathrm{fix}}
\newcommand{\opt}{\mathrm{opt}}
\newtheorem{restatedtheoremthree}{Theorem}

\begin{document}

\title{Fundamental Limits of Statistical Watermarking \\ for LLMs with Fixed and Optimizable \\ Common Randomness}

\author{%
    \IEEEauthorblockN{
    Haiyun~He\IEEEauthorrefmark{1}, 
    Yepeng~Liu\IEEEauthorrefmark{2}, 
    Zhuoer~Shen\IEEEauthorrefmark{2},
    Ziqiao~Wang\IEEEauthorrefmark{3},
    Yongyi~Mao\IEEEauthorrefmark{4}, 
    and Yuheng~Bu\IEEEauthorrefmark{2}}

   \IEEEauthorblockA{\IEEEauthorrefmark{1}%
                     {\small Internet of Things Thrust, HKUST (Guangzhou), Guangzhou, China}}

   \IEEEauthorblockA{\IEEEauthorrefmark{2}%
                     {\small Department of Computer Science, 
                     University of California, Santa Barbara, CA, USA}}

   \IEEEauthorblockA{\IEEEauthorrefmark{3}%
                     {\small School of Computer Science and Technology, 
                     Tongji University, Shanghai, China}}
    
   \IEEEauthorblockA{\IEEEauthorrefmark{4}%
                     {\small School of Electrical Engineering and Computer Science, 
                     University of Ottawa, Ottawa, ON, Canada}}
    
   \begin{footnotesize}
   Emails: haiyunhe@hkust-gz.edu.cn, 
   \{yepengliu, zhuoershen, buyuheng\}@ucsb.edu, 
   ziqiaowang@tongji.edu.cn, 
   ymao@uottawa.ca
   \end{footnotesize}
   \vspace{-1em}
}



\maketitle

\begin{abstract}
Watermarking has emerged as an important technique for distinguishing AI-generated text from human-written text. However, most existing work is primarily empirical, relies on ad hoc watermarking designs, and lacks a fundamental understanding of the performance limits. In this paper, we propose a unified theoretical framework for statistical watermarking of large language models (LLMs), casting the problem as an independence test between generated text and common randomness and jointly designing the watermarking embedder and the detector. Our objective is to maximize detection power while controlling the worst-case false positive rate (FPR) and limiting distortion to text quality. 
A central focus of our framework is the design of the common randomness, which plays a critical role in both watermark embedding and detection. We characterize the fundamental limits of statistical watermarking under both fixed and optimizable common randomness. Our results show that optimizing the common randomness does not improve the fundamental limits: the best achievable performance is the same as that with fixed common randomness. Nevertheless, common randomness design remains highly important because it can substantially simplify detector design. In particular, with optimizable common randomness, a fixed, deterministic, and model-agnostic detector can achieve optimal performance.
We further study a practical setting with fixed uniform common randomness and a fixed deterministic detector, and characterize its performance gap relative to the optimizable case, thereby demonstrating the practical value of optimizable common randomness.
\end{abstract}



\section{Introduction}\label{Sec: intro}

Large language models (LLMs) have greatly expanded the ability of machines to generate fluent and useful text, but they have also created urgent concerns about content provenance, misuse, plagiarism, and accountability \cite{touvron2023llama, liu2024survey, wu2025survey}. Watermarking has therefore emerged as an important tool for distinguishing AI-generated text from human-written text. Existing watermarking methods for AI-generated text are commonly divided into \emph{post-process} and \emph{in-process} approaches \cite{liu2024survey, wu2025survey}. Post-process watermarking is applied after text generation \cite{brassil1995electronic, yoo2023robust, yang2023watermarking, munyer2023deeptextmark, yang2022tracing, sato2023embarrassingly, zhang2024remark, abdelnabi2021adversarial, martinian2005authentication, moulin2000information, chen2000design, 1564436, 4418489}, whereas in-process watermarking embeds the watermark during generation \cite{wang2023towards, fairoze2023publicly, hu2023unbiased, huo2024token, zhang2023watermarks, tu2023waterbench, ren2023robust, hou2024semstamp, kirchenbauer2023watermark, zhao2023provable, liu2024adaptive, liu2024a, qu2024provably, fernandez2023three, fu2024gumbelsoft, wu2023dipmark, zhao2024permute, boroujeny2024multi, christ2024undetectable, giboulot2024watermax}. Among them, in-process watermarking is particularly attractive because it is invisible, flexible, and can be integrated into the generation with negligible latency. In this work, we focus exclusively on \emph{zero-bit watermarking}, where the objective is to detect the presence of a watermark rather than to decode an embedded message.

Despite this practical progress, the theoretical understanding of LLM watermarking remains limited. In practice, a useful watermarking scheme should achieve high detection power under a stringent false positive rate (FPR) constraint while preserving text quality. Many existing schemes do so heuristically, for example, by perturbing token logits as in the green-red-list watermark \cite{kirchenbauer2023watermark} or by modifying the sampling procedure as in Gumbel-Max-style methods \cite{gumbel2023}. Their detectors are typically built from handcrafted score statistics. Although these approaches can work well empirically, they do not answer a basic question: \emph{what is the fundamental limit of statistical watermarking under explicit constraints on distortion and FPR, and what watermarking schemes attain that limit?}

Information-theoretic tools have been widely used to characterize the fundamental limits of post-hoc watermarking, with important connections to state-dependent coding problems such as the Gel'fand--Pinsker problem and dirty-paper coding \cite{gelfand1980coding, costa1983writing}. In contrast, in-process watermarking for LLMs gives rise to a novel form of \emph{distributional information embedding} \cite{he2025distributional} that falls outside the scope of these classical results.

Recent theoretical works have taken important steps toward this question, but from different directions. \cite{takezawa2023necessary}, \cite{wouters2024optimizing}, and \cite{cai2024towards} analyze or optimize logit perturbation strategies for specific green-red-list-style schemes. \cite{huang2023towards} formulate watermarking as a statistical independence test between text and watermark and characterize the optimal watermarking scheme for a \emph{fixed detector}. 
In contrast, \cite{li2024statistical} optimize the detector for a \emph{fixed watermarking scheme} via i.i.d.\ pivotal statistics. 
Recently, a closely related work \cite{long2025optimized} studies coupling-based LLM watermarking, where a randomized vocabulary partition is optimized and used in both watermark generation and detection to improve the induced coupling between fixed side information and the next-token distribution under a worst-case min-entropy constraint. It focuses on a one-shot/token-level setting and adopts a Bayesian testing formulation. Similarly, \cite{tsurheavywater} develops watermarking methods by formulating token-level watermark design as an optimal-transport-based coupling problem with fixed common randomness, an optimizable score function, and a detection-gap objective, especially for low-entropy next-token distributions.
These works provide valuable insights, but they do not give a unified characterization of the \emph{fundamental limit} together with the corresponding \emph{optimal watermarking scheme design} under a general framework.

This paper develops such a framework by focusing on the role of \emph{common randomness}. In our formulation, watermarking is cast as an independence testing problem between the generated text and a shared random sequence available to both the watermark embedder and the detector. This viewpoint is closely related to classical themes in information and coding theory, where common randomness plays a central role in random coding for channel coding \cite{cover2006elements}, shared secret keys for secrecy systems \cite{shannon1949communication}, and common-randomness-assisted coding for arbitrarily varying channels (AVCs) \cite{blackwell1960capacities, ahlswede1978elimination}. In these settings, the common randomness is usually treated as fixed as part of the problem. In LLM watermarking, however, it is natural to ask a broader question: should the common randomness be treated as fixed, or can its distribution itself be optimized as part of the watermark design? We assume that the common randomness is available to both the embedder and the detector, without imposing a separate constraint on its entropy or the number of shared random bits per token. The distinction between fixed and optimizable common randomness concerns whether its distribution is prescribed or designed, rather than a constraint on its rate.

This distinction leads to the main perspective of the paper. We study statistical watermarking under a general framework with either fixed or optimizable common randomness, and organize the theory through a \emph{fundamental limits (converse) vs.\  optimal watermarking design (achievability)} lens. 
On one hand, the converse identifies the fundamental limit, namely the minimum achievable Type-II error under a worst-case Type-I error constraint with a distortion budget. On the other hand, the achievability results provide optimal watermarking designs, namely jointly optimal watermark embedders and detectors that attain this limit under different common-randomness settings. This perspective clarifies the gap between theory and practice: the lower bound specifies what no scheme can beat, while the constructions show how optimal schemes look like.

Our results show that optimizing the common randomness does \emph{not} improve the fundamental limit attainable with fixed continuous common randomness. The same optimal Type-II error is also achievable with fixed discrete common randomness satisfying the conditions in Theorem~\ref{thm:randomized-discrete-optimal-construction}. Thus, from a fundamental-limits viewpoint, optimizable common randomness does not enlarge what is achievable with a suitably chosen fixed common randomness. 
Its benefit lies elsewhere: it simplifies optimal watermarking design. In particular, under optimizable common randomness, a fixed deterministic and model-agnostic (i.e., independent of the original LLM generation distribution) detector can achieve the optimum. 
We also analyze a practically motivated setting with fixed uniform common randomness and a fixed deterministic model-agnostic detector, and characterize its performance gap relative to the optimizable benchmark. We emphasize that the achievability results in this paper are primarily sequence-level and existential: they characterize what is information-theoretically attainable under the stated statistical watermarking framework. Direct implementation in autoregressive LLM generation requires token-wise adaptation and a practical mechanism for generating common randomness. Our conference paper \cite{he2025theoretically} follows this direction by applying the theory-guided optimal embedding rule to each next-token-prediction distribution and aggregating token-wise detection scores.

It is also worth noting the relationship between the present paper and our conference paper \cite{he2025theoretically}. The conference version studies a specific watermarking problem with optimizable common randomness and proceeds toward practical algorithm design. This paper plays a broader and more foundational role: it studies the fundamental limits of statistical watermarking under a general framework that includes both fixed and optimizable common randomness, and interprets the achievability results as optimal watermarking designs. In this sense, the two papers are complementary: this paper develops the general theory, whereas the conference version focuses on one practically motivated instance together with an implementable algorithm.
Furthermore, in contrast to the closely related token-level formulations \cite{long2025optimized,tsurheavywater}, which use either a Bayesian testing criterion or a detection-gap objective under fixed side-information distributions, our work characterizes sequence-level Neyman--Pearson fundamental limits under worst-case Type-I error and distortion constraints, including the optimizable-common-randomness setting.

\begin{table*}[t!]
\small
\centering
\caption{Summary of the fundamental limits of optimal watermarking schemes under different settings. Although the converse results are identical in all cases, the achievability results lead to different detector structures, and in practice, a deterministic, model-agnostic detector is preferred. 
}
\label{tab:summary}

\begin{tabular}{l|c|c|c}
\toprule
\textbf{\makecell{Common \\ Randomness}} & \textbf{Converse} & \textbf{Detector} & \textbf{Achievable} \\
\midrule

\makecell{Fixed continuous \\ (Thm.~\ref{thm:continuous-optimal-construction})}
& \multirow{8}{*}{\makecell{$\min\limits_{\substack{P_{X^T}: \sD(P_{X^T},Q_{X^T})\leq \epsilon}} \sum\nolimits_{x^T}(P_{X^T}(x^T)-\alpha)_+$}}
& \makecell{Deterministic \\ Model-aware}
& Yes \\
\cmidrule(lr){1-1}\cmidrule(lr){3-4}

\makecell{Fixed discrete \\ (Thm.~\ref{thm:randomized-discrete-optimal-construction})}
& 
& \makecell{Randomized \\ Model-aware}
& \makecell{Not always; \\
necessary and sufficient \\
conditions established}\\
\cmidrule(lr){1-1}\cmidrule(lr){3-4}


\makecell{Optimizable \\ (Thm.~\ref{Thm: optimal detector})}
&
& \makecell{Deterministic \\ Model-agnostic} 
& Yes \\

\bottomrule
\end{tabular}
\end{table*}

The main contributions of this paper are summarized as follows.
\begin{itemize}
    \item We propose a unified information-theoretic framework for statistical watermarking of LLMs by formulating watermark detection as an independence test between generated text and common randomness.
    \item We characterize the fundamental limit of statistical watermarking under distortion and worst-case FPR constraints, for both fixed and optimizable common randomness.
    \item As summarized in Table \ref{tab:summary}, optimizing the common randomness does not improve the fundamental limit (converse): fixed and optimizable common randomness both lead to the same best achievable Type-II error.
    \item We construct jointly optimal watermark embedders and detectors under multiple common-randomness settings, thereby identifying the corresponding optimal watermarking designs that attain the fundamental limit.
    \item We show that, although optimizable common randomness does not improve the limit itself, it can substantially simplify detector design by enabling a fixed deterministic and model-agnostic optimal detector.
    \item We analyze a practical setting with fixed uniform common randomness and a fixed, model-agnostic deterministic detector, and quantify its performance gap to the optimizable benchmark that jointly optimizes the watermark embedder, detector and common randomness.
\end{itemize}

\section{Problem Formulation}

\subsection{Notation} 
For any set $\cX$, we denote the space of all probability measures over  $\cX$ by $\cP(\cX)$. 
For a sequence of random variables $X_1, \ldots, X_n$, and any $i, j \in [n]$ with  $i\le j$, we denote $X_i^j\coloneqq (X_i,\ldots,X_j)$ and suppress the subscript when $i=1$. 
We may use distortion measure, namely, a function $\sD: \calP(\cX) \times \calP(\cX) \to [0,+\infty)$
to measure the dissimilarity between two distributions in $\calP(\cX)$. 
For example, 
the total variation distance, as a distortion, between $\mu,\nu\in\cP(\cX)$ is $\sD_\TV(\mu,\nu)\coloneqq\int \frac{1}{2}| \frac{d  \mu}{ d  \nu}-1| \,  d \nu$. 
For any set $A\subseteq \cX$, we use $\delta_A$ to denote its indicator function, namely, $\delta_A(x)\coloneqq \mathbbm{1}\{x\in A\}$. Additionally, we denote $(x)_+:=\max\{x, 0\}$ and $x\wedge y:= \min\{x, y\}$.

\subsection{Statistical Watermarking for LLMs}
\paragraph{Tokenization and NTP in LLMs}
LLMs process text through ``tokenization,'' namely, breaking it down into words or word fragments called ``tokens''. An LLM generates text token-by-token, and different LLM families use different tokenization methods.
Let $\calV$ denote the token vocabulary, typically of size $|\calV|=\calO(10^4)$ \cite{liu2019roberta,radford2019language,zhang2022opt,touvron2023llama}.
An \emph{unwatermarked} LLM generates the next token $X_t$ based on a prompt $\pt$ and the previous tokens $x^{t-1}$
by sampling the Next-Token Prediction (NTP) distribution $Q_{X_t|x^{t-1},\pt}$. For simplicity, the prompt dependency is suppressed in notation throughout the paper. The joint distribution of a length-$T$ generated token sequence $X^T$ is then given by $Q_{X^{T}}\coloneqq\prod_{t=1}^T Q_{X_t|X^{t-1}}$.

\paragraph{Distributional Watermark Embedding with Common Randomness}
Unlike traditional post-hoc detection methods that rely on partitioning the text space by assuming certain sentences are unlikely to be human-written, modern LLM watermarking schemes instead embed a watermark into the generation distribution of LLM and leverage the statistical dependence between text $X^T$ and an auxiliary random sequence $\zeta^T$ to overcome this restriction.

\begin{figure}[!t]
\vspace{-1.5em}
\centering
\begin{tikzpicture}[scale=0.7,every node/.style={transform shape}]



    \node (start) at (-1, 4.5) {};

    \tikzset{
        mybox/.style={
            rectangle, 
            draw, 
            fill=white, 
            minimum height=1cm, 
            minimum width=2cm, 
            align=center,
            thick,
            drop shadow
        },
        myboxtall/.style={
            rectangle, 
            draw, 
            fill=white, 
            minimum height=1.5cm, 
            minimum width=2cm, 
            align=center,
            thick,
            drop shadow
        },
        cloudnode/.style={cloud, draw, fill=white, cloud puffs=10, cloud puff arc=120, aspect=4, inner ysep=1em, minimum width=1cm, minimum height=.5cm},
        roundnode/.style={ellipse, draw, fill=white, minimum width=2cm, minimum height=1cm}
    }

        \node[myboxtall] (sampler1) at (9.5, 4.5) {Sampler};
        \node[mybox] (WM) at (5, 4) {Watermark \\ Embedder};
        \node[myboxtall, below = 2 of WM] (dec) {Watermark \\ Detector};
        \node[] (common) at (5,2.5) {Common Randomness $\zeta^T\sim P_{\zeta^T}$};

        \node[gray] at (0.2,2.8) {Watermark Generation};
        \draw[-,dashed, thick, gray] (-1.6,2.5)--++(3.5,0);
        \node[gray] at (0.2,2.3) {Watermark Detection};

        \node[myboxtall] (LLM) at (0.5,4.5) {\Large \faDesktop \\ LLM};
        \draw[->, dashed, very thick] (LLM.east)++(0, 0.5) -- (8.5, 5);
        \node[] at (7,5.3) {$Q_{X^T}$};
        \draw[->, dashed, very thick] (sampler1.south)++(0.5,0) -- ++(0, -3.5) -- ++(-3.8, 0) node[midway, above] {$x^T$} node[midway, below] {unwatermarked text};

        \draw[->, very thick, color = SteelBlue2] (LLM.east)++(0, -0.5) -- (WM.west) node[near end, above] {$Q_{X^T}$};
        \draw[->, very thick, color = SteelBlue2] (WM.east) -- (8.5,4) node[midway, above] {$P_{X^T|\zeta^T}$};
        \draw[->, very thick, color = SteelBlue2] (sampler1.south)++(-.5,0) -- ++(0,-2.5) -- ++(-2.8,0) node[midway, above] {$x^T$} node[midway, below] {watermarked text};

        \draw[->, densely dotted, very thick, color = gray] (common.north) -- (WM.south);
        \draw[->, densely dotted, very thick, color=gray] (common.south) -- (dec.north);
        
        \draw[->, very thick, color = SteelBlue2] (dec.west)++(0, 0.5) -- ++(-2,0) node[midway, above] {1};
        \draw[->, dashed, very thick] (dec.west)++(0, -0.5) -- ++(-2,0) node[midway, above] {0};
        \node at (0.3,1.2) {Watermarked};
        \node at (0.3,.2) {Unwatermarked};

\end{tikzpicture}
    \caption{Overview of LLM watermarking and detection.}
    \label{Fig: system}
    \vspace{-1em}
\end{figure}
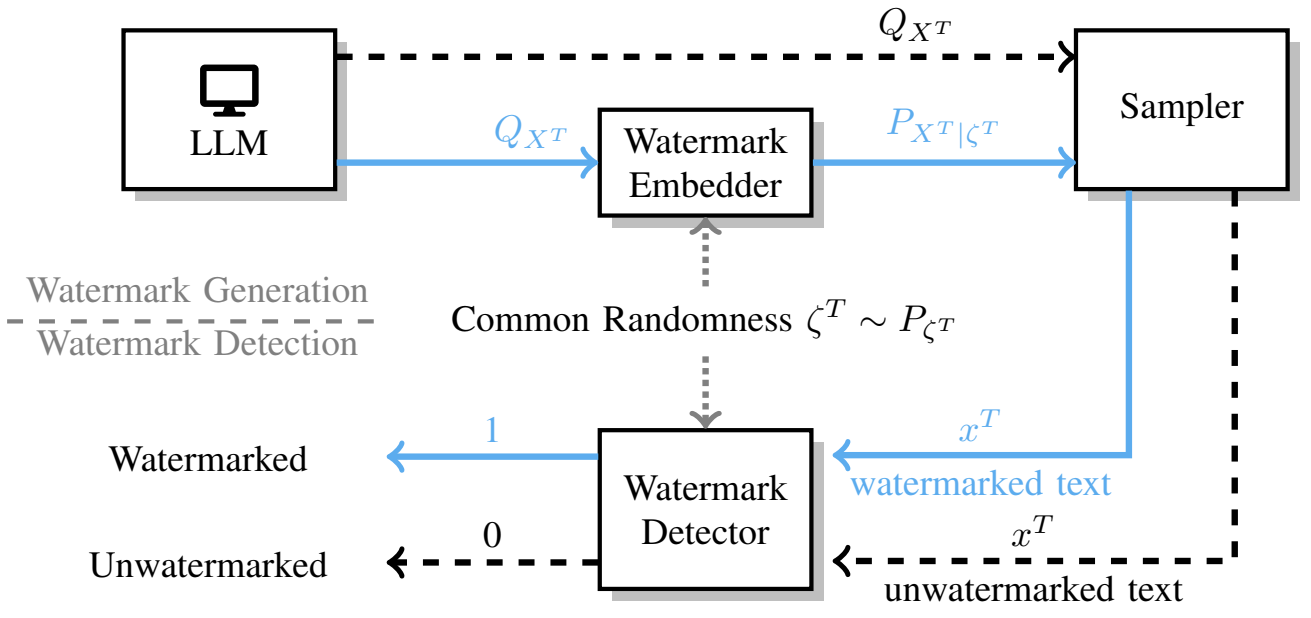

In this paper, we consider a general formulation of statistical watermarking for LLMs, as shown in Figure \ref{Fig: system}, which encompasses most of the existing watermarking schemes (discussed in \cite[Appendix B]{he2025theoretically}). The watermark embedder and detector share a \emph{common randomness} represented by the auxiliary random sequence $\zeta^T$ drawn from a space $\calZ^T$ (either discrete or continuous). After passing through a watermark embedder, the \emph{watermarked} LLM samples the next token according to the \emph{modified} NTP distribution $P_{X_t|x^{t-1},\zeta^T}$, where  $P_{X^T|\zeta^T}=\prod_{t=1}^T P_{X_t|X^{t-1},\zeta^T}$.  This process thus associates the watermarked token sequence $X^T$ with the auxiliary sequence $\zeta^T$ statistically, enabling the detector to distinguish whether a received sequence $X^T$ is watermarked or not based on the common randomness.

Specifically, our formulation allows the common random sequence $\zeta^T$ to take an arbitrary structure, not necessarily i.i.d. One core question we aim to address is how watermark design depends on the structure of $P_{\zeta^T}$, for example, whether it is discrete or continuous and whether the distribution $P_{\zeta^T}$ is fixed or optimized.  In practice, the common randomness is shared via a secret key accessible during both watermark embedding and detection.

Note that the resulting joint distribution of the watermarked token sequence $X^T$, denoted by $P_{X^T}$, might be different from the original $Q_{X^T}$. To evaluate the \emph{distortion level} of a watermarking scheme, we measure the statistical divergence between  $P_{X^T}$ and $Q_{X^T}$.
\begin{definition}[$\epsilon$-distorted watermarking]
    A watermarking scheme is \emph{$\epsilon$-distorted} with respect to distortion $\sD$, if $\sD(P_{X^T}, Q_{X^T})\leq \epsilon$. Here, $\sD$ can be any distortion metric.
\end{definition}
 Common examples of such divergences include total variation, squared distance, KL divergence, and Wasserstein distance. If $\epsilon=0$, the watermarking scheme is called \emph{distortion-free}.

\paragraph{Watermark Detection with Common Randomness}
With the shared common randomness, the detector has access to the auxiliary sequence $\zeta^T$ and receives a token sequence $X^T$. When $X^T$ is generated by a watermarked LLM, it depends on $\zeta^T$ statistically, while human-generated $X^T$ and $\zeta^T$ are independent.
Therefore, given a pair of sequences $(X^T,\zeta^T)$, the detection task boils down to discriminating between the following two hypotheses:
\begin{itemize}[leftmargin=*]
    \item $\rmH_0$: $X^T$ is generated by a human/unwatermarked LLM, i.e., $(X^T,\zeta^T)\sim Q_{X^T}\otimes P_{\zeta^T}$;
    \item $\rmH_1$: $X^T$ is generated by a watermarked LLM, i.e., $(X^T,\zeta^T)\sim P_{X^T,\zeta^T}$.
\end{itemize}
We consider a detector $\gamma: \calV^{T}\times \calZ^T \to \{0,1\}$, which maps $(X^T,\zeta^T)$ to the index of one of the two hypotheses.  
Detection performance is measured by  Type-I (false-alarm) and Type-II (miss-detection) error probabilities:
\begin{align}
    \text{Type-I:}\quad &\beta_0(\gamma,Q_{X^T},P_{\zeta^T}):=\bbE_{Q_{X^T},P_{\zeta^T}}[\gamma(X^T,\zeta^T)], \label{Def: type I}\\
    \text{Type-II:}\quad &\beta_1(\gamma,P_{X^T,\zeta^T} ):=\bbE_{P_{X^T,\zeta^T}}[1-\gamma(X^T,\zeta^T)].
    \label{Def: type II}
\end{align}
When $Q_{X^T}$, $P_{\zeta^T}$ and $P_{X^T,\zeta^T}$ are fixed, it is well-known that the optimal detector is a likelihood-ratio test \cite{thomas2006elements}. However, in watermark detection, the null distribution $Q_{X^T}$ may correspond to arbitrary human-written text with diverse and unknown structures. As a result, we must account for controlling the Type-I error over \emph{all possible} $Q_{X^T}$.
Furthermore, in our formulation,  the watermark-induced distribution $P_{X^T|\zeta^T}$ is a design parameter and remains unknown to the detector. The common randomness $P_{\zeta^T}$ may either be fixed a priori or optimized depending on the scenario. 
Consequently, the optimal detector is no longer given by a standard likelihood-ratio test. This motivates the following optimization-based formulation under worst-case conditions.

\paragraph{Optimization Problem}
Our goal is to jointly design an $\epsilon$-distorted watermarking scheme and a detector that minimizes the Type-II error while ensuring the \emph{worst-case} Type-I error  $\sup_{Q_{X^T}}\beta_0(\gamma,Q_{X^T},P_{\zeta^T})$ under a constant $\alpha\in(0,1)$. 

We consider two settings: one where the common randomness distribution $P_{\zeta^T}$ is fixed, and another where it is treated as an optimization variable. The corresponding optimization problem is:
\begin{enumerate}
    \item Fixed common randomness:
    \begin{align}
       &\inf_{\gamma,P_{X^T|\zeta^T}} \beta_1(\gamma,P_{X^T,\zeta^T} ) \tag{P1} \label{Eq: opt-fixed}\\
       &\text{ s.t. } \sup_{Q_{X^T}}\beta_0(\gamma,Q_{X^T},P_{\zeta^T})\leq \alpha,\; \sD(P_{X^T}, Q_{X^T})\leq \epsilon. \nn 
    \end{align}
    \item Optimizable common randomness:
    \begin{align}
       &\inf_{\gamma,P_{X^T,\zeta^T}} \beta_1(\gamma,P_{X^T,\zeta^T} ) \tag{P2} \label{Eq: opt-optimizable}\\
       &\text{ s.t. } \sup_{Q_{X^T}}\beta_0(\gamma,Q_{X^T},P_{\zeta^T})\leq \alpha,\; \sD(P_{X^T}, Q_{X^T})\leq \epsilon.  \nn
    \end{align}
\end{enumerate}
The corresponding optimal objective value is the \emph{minimum Type-II error} under the given framework, denoted by $\beta_{1,\fix}^*(Q_{X^T}, P_{\zeta^T}, \alpha, \epsilon)$ and $\beta_{1,\opt}^*(Q_{X^T}, \alpha, \epsilon)$, respectively.

\section{Optimal Watermarking Scheme with Fixed Common Randomness}\label{Sec:opt fixed common}
In this section, we aim to solve the optimization in \eqref{Eq: opt-fixed} and identify the jointly optimal watermark embedder and detector. However, solving \eqref{Eq: opt-fixed} directly is challenging due to the binary nature of $\gamma$ and the vast set of possible detectors $\gamma$, sized $2^{|\calV|^T|\calZ|^T}$. To address this, we begin with a fixed $\gamma$ with an arbitrary acceptance region and derive a lower bound for the minimum Type-II error of \eqref{Eq: opt-fixed}, which holds for any detector $\gamma$. 
Later, we show that this lower bound is achievable by a deterministic detector when $P_{\zeta^T}$ is continuous, and by a randomized detector when $P_{\zeta^T}$ is discrete. Therefore, it characterizes the fundamental limit of the minimum Type-II error in \eqref{Eq: opt-fixed}.


\begin{lemma}[Fundamental limit of the Minimum Type-II Error under Fixed Common Randomness]\label{Thm:min typeII error fixed}
     Given any LLM with generation distribution $Q_{X^T}$, any fixed auxiliary distribution $P_{\zeta^T}$,  $\alpha\in[0,1]$ and $\epsilon\geq 0$,  the universally minimum Type-II error attained from \eqref{Eq: opt-fixed} is lower bounded by 
    \begin{equation}
     \beta_{1,\fix}^*(Q_{X^T}, P_{\zeta^T}, \alpha, \epsilon)\geq \!\!\! \min_{\substack{P_{X^T}: \\ \sD(P_{X^T}, Q_{X^T} )\leq \epsilon}}\sum\nolimits_{x^T}(P_{X^T}(x^T)- \alpha)_+.  \label{Eq: Type-II LB}
    \end{equation}
     Moreover, if the distortion measure is chosen as the total variation distance $\rvD_\TV(P_{X^T},Q_{X^T})$,  the lower bound admits the closed-form expression
    \begin{align}
        &\beta_{1,\fix}^*(Q_{X^T}, P_{\zeta^T}, \alpha, \epsilon) \nn\\
        &\geq\!\!\!\bigg(\sum_{x^T}\left(Q_{X^T}(x^T)- \alpha\right)_+ \!\!\!- \!\Big(\epsilon \wedge \sum_{x^T}\big( \alpha-Q_{X^T}(x^T)\big)_+ \Big)\bigg)_+.
    \end{align}
\end{lemma}

The proof of Lemma \ref{Thm:min typeII error fixed} is deferred to Supplementary~\ref{App:converse min error fixed}. We note that the lower bound in \eqref{Eq: Type-II LB} is maximized at the watermarked text distribution
 \begin{equation}
    P_{X^T}^*\coloneqq\argmin_{P_{X^T}:\sD(P_{X^T}, Q_{X^T} )\leq \epsilon}\sum\nolimits_{x^T}(P_{X^T}(x^T)- \alpha)_+,  \label{Eq: opt X marginal}
    \end{equation}
which is equal to $Q_{X^T}$ when $\epsilon=0$.

\begin{wrapfigure}[11]{r}{.6\linewidth}
    \vspace{-20pt}
    \begin{center}
    \includegraphics[trim={2.5cm .4cm .4cm 0}, width=.95\linewidth]{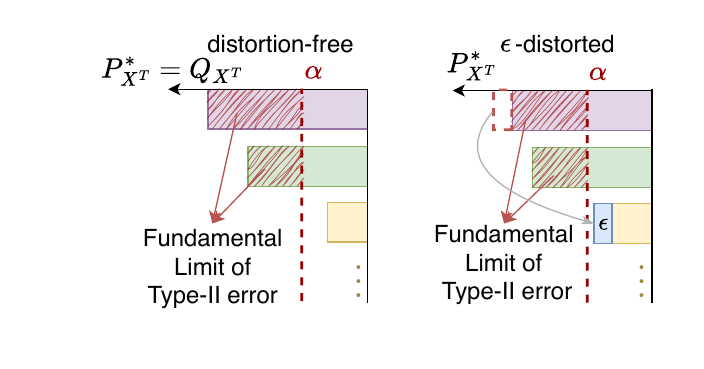}
    \vspace{-5pt}
    \caption{Illustration of error--distortion trade-off when $\epsilon < \sum_{x^T}( \alpha-Q_{X^T}(x^T))_+$.}
    \label{Fig: marginal}
    \end{center}
\end{wrapfigure}    
Lemma \ref{Thm:min typeII error fixed} shows that, for any watermarking scheme with fixed common randomness, the fundamental limit of detection performance depends on the original NTP distribution of the LLM. When the original $Q_{X^T}$ is more concentrated (low entropy), the minimum achievable Type-II error increases. This hints that it is inherently difficult to watermark low-entropy text. However, increasing the allowable distortion 
$\epsilon$ can enhance the capacity for reducing detection errors, as illustrated in Figure \ref{Fig: marginal}. Throughout the paper, we use the term \emph{optimal watermarking scheme} to refer to an embedder–detector pair that minimizes the objective in \eqref{Eq: opt-fixed} (or \eqref{Eq: opt-optimizable}), with the resulting minimum value attaining the lower bound in Lemma \ref{Thm:min typeII error fixed}.

Next, we present constructions of jointly optimal watermark embedders and detectors that attain the lower bound in Lemma \ref{Thm:min typeII error fixed} under different choices of the common randomness $P_{\zeta^T}$. We begin by distinguishing between continuous and discrete $P_{\zeta^T}$. For continuous $P_{\zeta^T}$, we establish that an optimal construction always exists under deterministic detectors. In contrast, when $P_{\zeta^T}$ is discrete, by relaxing the detector from deterministic to randomized, we show that an optimal construction can be achieved for arbitrary discrete $P_{\zeta^T}$. When the detector is restricted to be deterministic, optimality is guaranteed only for a special case of uniform distributions and may fail for general discrete distributions.

Before presenting the theorem, we define a collection of important quantities that will be used throughout the analysis:
\begin{equation}
m(x^T)\coloneqq P^*_{X^T}(x^T)\wedge \alpha, \qquad \forall x^T\in\calV^T, \label{Eq: m(x)}
\end{equation}
where $P^*_{X^T}$ is given in \eqref{Eq: opt X marginal}. Intuitively, the quantities $\{m(x^T)\}$ characterize the desired per-sequence contribution to the detection power, with each contribution capped so as to respect the worst-case Type-I error constraint $\alpha$. Note that the worst-case Type-I error also covers the case where $Q_{X^T}$ is a point mass. In optimization problem \eqref{Eq: opt-fixed}, we notice that the lower bound in Lemma \ref{Thm:min typeII error fixed} can be achieved if and only if the watermark embedder-detector pair satisfies $\bbE_{P_{\zeta^T}}[P_{X^T|\zeta^T}(x^T|\zeta^T)\gamma(x^T,\zeta^T)]=m(x^T)$ for all $x^T$. Equivalently, the constraint budget must be fully utilized at every $x^T$ in order to attain the minimum objective value; see \eqref{Eq: 1-TypeII error}--\eqref{Eq: 1-TypeII error ub} in Supplementary~\ref{App:converse min error fixed}. Therefore, the achievability constructions presented below critically depend on the structure of $\{m(x^T)\}$.

\subsection{Continuous Common Randomness}
We first consider the case where $P_{\zeta^T}$ is continuous and the detector $\gamma: \calV^{T}\times \calZ^T \to \{0,1\}$ is deterministic. In this setting, the following result establishes that an optimal construction always exists. The key takeaway in the optimal design is: (i) the deterministic detector $\gamma$ composing the jointly optimal construction depends on the distribution $P_{\zeta^T}$; (ii) the watermark embedder outputs a joint distribution belonging to the coupling set $\Pi(P^*_{X^T},P_{\zeta^T})$ by solving an optimal transport problem.

\begin{theorem}[Continuous common randomness: jointly optimal watermark embedder and deterministic detector]
\label{thm:continuous-optimal-construction}
Fix an LLM with generation distribution $Q_{X^T}$, $\alpha \in [0,1]$ and $\epsilon>0$. 
Let $P_{\zeta^T}$ be a nonatomic probability distribution on a continuous space $\mathcal Z^T$.
There exists a collection of pairwise disjoint measurable sets $\{B(x^T)\}_{x^T\in\calV^T}\subseteq \calZ^T$ such that
$P_{\zeta^T}(B(x^T))=m(x^T), \  \forall x^T\in\calV^T$.

Denote $B_0:=\mathcal Z^T\setminus \bigcup_{x^T\in\calV^T} B(x^T)$, called the redundant bucket, and $P_{\zeta^T}(B_0)=1-\sum_{x^T}m(x^T)$.
Then the jointly optimal construction that minimizes the Type-II error in \eqref{Eq: opt-fixed} and attains the lower bound in Lemma \ref{Thm:min typeII error fixed} is:
\begin{enumerate}[leftmargin=*]
    \item deterministic detector $\gamma^*(X^T,\zeta^T)=\mathbbm{1}\{\zeta^T\in B(X^T)\}$;
    \item watermark embedder with output coupling $P^*_{X^T,\zeta^T}$: for each $x^T\in\calV^T$, and for any measurable set $A \subseteq \calZ^T$,
    \begin{align}
        &\!\!\!\!\!\! P^*_{X^T,\zeta^T}(x^T,A)
        \\
        &\!\!\!\!\!\! =\begin{cases}
            \begin{aligned}[t]
                &P_{\zeta^T}(A\cap B(x^T))\\
                &\quad{}+
                \dfrac{P^*_{X^T}(x^T)-m(x^T)}
                      {P_{\zeta^T}(B_0)}
                P_{\zeta^T}(A\cap B_0),
            \end{aligned}
            & \!\!\!\! P_{\zeta^T}(B_0)>0,\\
            P_{\zeta^T}(A\cap B(x^T)),
            & \!\!\!\! P_{\zeta^T}(B_0)=0.
        \end{cases}
    \end{align}
\end{enumerate}
\end{theorem}

\begin{figure*}
    \centering
    \includegraphics[width=0.82\linewidth]{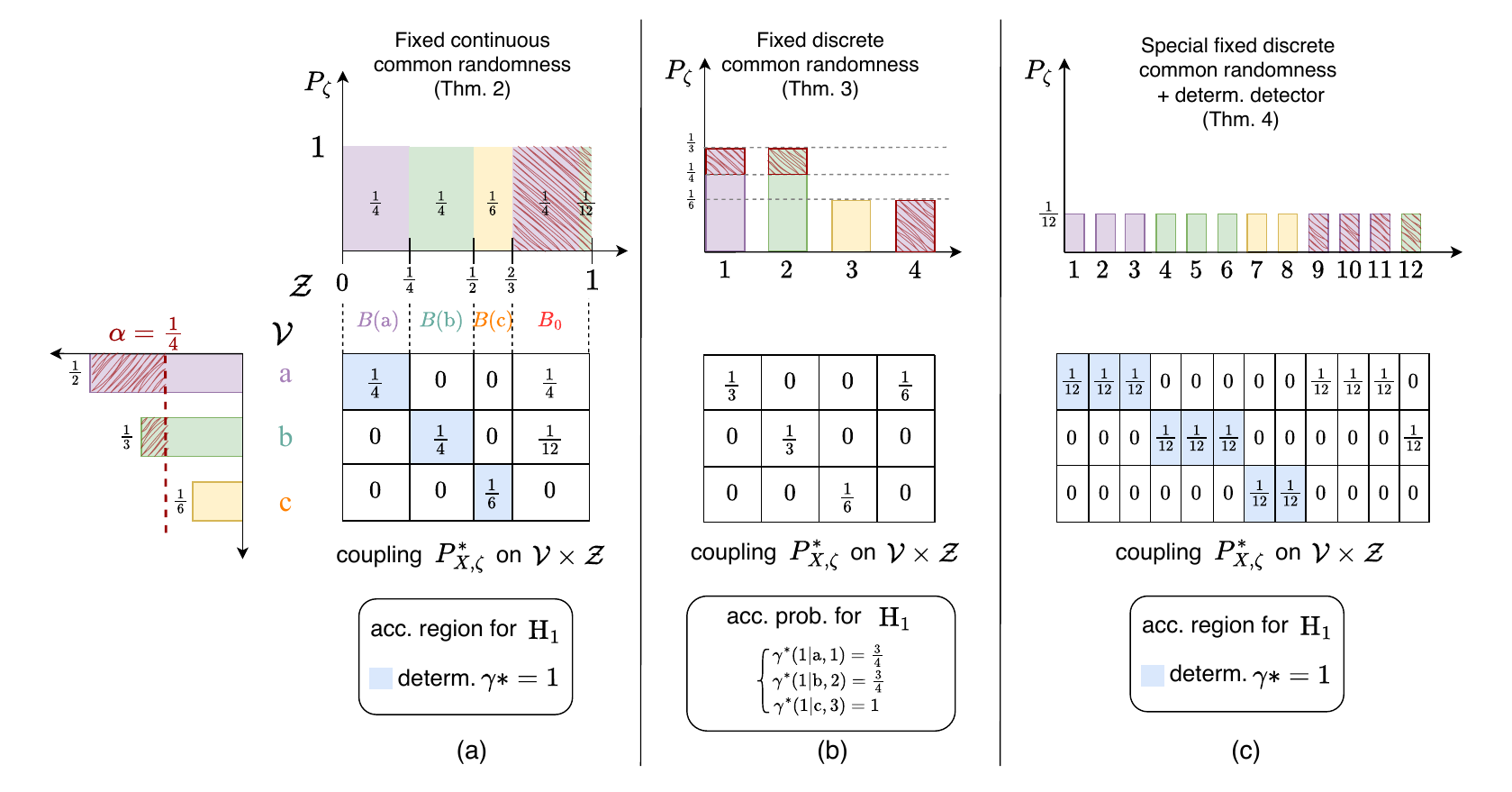}
    \vspace{-5pt}
    \caption{Toy examples of the optimal detector and coupling under fixed common randomness when $T=1$ and $\calV=\{\text{a},\text{b},\text{c}\}$.}
    \label{fig:fixed CR example}
\end{figure*}

The proof of Theorem \ref{thm:continuous-optimal-construction} is deferred to Appendix \ref{App: pf of thm:cont-opt-construction}. Note that the the second term in the first line of $P^*_{X^T,\zeta^T}(x^T,A)$ is equal to $0$ if $P_{\zeta^T}(B_0)=0$, as there will be no redundant mass and $P^*_{X^T}(x^T)=m(x^T)$ for all $x^T$. To better illustrate the achievability construction, here is a toy example when $T=1$, which can be directly extended to any $T$. Consider a fixed uniform common randomness $P_\zeta=\Unif([0,1])$, a token vocabulary $\calV=\{\text{a,b,c}\}$, a watermarked distribution $P_X^*=(\frac{1}{2},\frac{1}{3},\frac{1}{6})$ and a false-alarm constraint $\alpha=\frac{1}{4}$. The resulting quantities in \eqref{Eq: m(x)} are $m(\text{a})=\frac{1}{4},m(\text{b})=\frac{1}{4},m(\text{c})=\frac{1}{6}$.  By setting $B(\text{a})=[0,\frac{1}{4}],B(\text{b})=(\frac{1}{4},\frac{1}{2}], B(\text{c})=(\frac{1}{2},\frac{2}{3}]$ and the redundant bucket $B_0=(\frac{2}{3},1]$, the optimal coupling $P^*_{X,\zeta}$ is constructed by moving $m(x)$ amount of mass from $P_X^*(x)$ to $P_{\zeta}(B(x))$, which is shown as a table in column (a) of Figure \ref{fig:fixed CR example}. The corresponding acceptance region for $\rmH_1$ is $\{(x,B(x))\}_{x\in\calV}$, shadowed by blue in Figure \ref{fig:fixed CR example}.
Several remarks are as follows.


\textbf{(i) Measure-preserving partition and detector structure.} 
For continuous $P_{\zeta^T}$, it is always possible to partition $\mathcal Z^T$ into a collection of pairwise disjoint measurable subsets whose probabilities exactly match $\{m(x^T)\}_{x^T\in\mathcal V^T}$, together with a remaining set that collects the leftover mass. This naturally yields the optimal detector $\gamma^*$, which maps each token sequence $x^T$ to a unique subset $B(x^T)\subseteq \mathcal Z^T$.

\textbf{(ii) Redundant bucket for worst-case Type-I error control.} To control the worst-case Type-I error, the construction of $P_{X^T,\zeta^T}^*$ uses a redundant bucket $B_0\subseteq \mathcal Z^T$ to absorb the excess mass of $P_{X^T}^*$ above $\alpha$. Whenever $\zeta^T\in B_0$, the detector $\gamma^*$ always declares the observed token sequence to be unwatermarked. In particular, if $P_{X^T}^*(x^T)>\alpha$, corresponding to a low-entropy text sequence $x^T$ such as a celebrity’s name, then part of the mass associated with $x^T$ must be assigned to $B_0$, making such sequences harder to detect as watermarked. Therefore, the optimal coupling $P_{X^T,\zeta^T}^*$ effectively controls the false positive rate for low-entropy texts.

\textbf{(iii) Model-awareness and connection to inverse transform sampling.} Both the optimal detector $\gamma^*$ and the optimal coupling $P_{X^T,\zeta^T}^*$  are \emph{model-aware}, depending on the original LLM distribution $Q_{X^T}$, which induces $P_{X^T}^*$ in \eqref{Eq: opt X marginal}, as well as on the prescribed common randomness distribution $P_{\zeta^T}$. In the extreme case where $P_{X^T}^*(x^T)\le \alpha$ for all $x^T$, the optimal construction reduces to a deterministic coupling induced by a measure-preserving partition of $\mathcal Z^T$, which can be viewed as a \emph{generalized inverse transform sampling scheme}. In particular, when $P_{\zeta^T}$ is uniform on $[0,1]^T$, the inverse transform watermark embedding rule in \cite{kuditipudi2023robust} coincides with the optimal coupling $P_{X^T,\zeta^T}^*$. However, their detector does not assign each token sequence $x^T$ to a unique subset of the auxiliary space, in contrast to the optimal detector $\gamma^*$ as in Theorem~\ref{thm:continuous-optimal-construction}. Consequently, their watermarking scheme does not attain the fundamental limit of the minimum Type-II error.

\subsection{Discrete Common Randomness}
In contrast to the case of continuous common randomness $P_{\zeta^T}$, where one can always construct disjoint subsets $B(x^T)\subseteq \calZ^T$ satisfying $P_{\zeta^T}(B(x^T))=m(x^T)$, the main difficulty in attaining the minimum Type-II error lower bound in Lemma \ref{Thm:min typeII error fixed} for an arbitrarily fixed discrete $P_{\zeta^T}$ stems from the inability to split discrete atoms $\zeta^T$.
Allowing randomized detectors provides additional flexibility in addressing this limitation. Specifically, instead of a deterministic rule $\gamma:\calV^T\times\calZ^T\to\{0,1\}$, we allow $\gamma(\cdot | x^T,\zeta^T)\in\calP(\{0,1\})$, so that the detector declares the observed pair $(x^T,\zeta^T)$ to be watermarked with probability $\gamma(1 | x^T,\zeta^T)$. Such partial acceptance can enable attainment of the lower bound when deterministic detection cannot, but it does not guarantee achievability for an arbitrary fixed discrete $P_{\zeta^T}$. We therefore first establish necessary and sufficient conditions on $P_{\zeta^T}$, for a given text distribution $P_{X^T}^*$ and false-alarm threshold $\alpha$, under which a coupling
and a randomized detector jointly achieve the lower bound. We then identify a class of discrete uniform common-randomness distributions for which the lower bound remains attainable with a deterministic detector.

\paragraph{Randomized Detector}
In the setting with a fixed discrete common randomness and a randomized detector,
the Type-I and Type-II error probabilities are rewritten as
\begin{align}
    \text{Type-I:}&\quad \beta_0(\gamma,Q_{X^T},P_{\zeta^T})=\bbE_{Q_{X^T},P_{\zeta^T}}[\gamma(1|X^T,\zeta^T)], \label{Def: type I soft}\\
    \text{Type-II:}&\quad \beta_1(\gamma,P_{X^T,\zeta^T} )=\bbE_{P_{X^T,\zeta^T}}[1-\gamma(1|X^T,\zeta^T)],
    \label{Def: type II soft}
\end{align}
which are essentially the same as \eqref{Def: type I} and \eqref{Def: type II}, with $\gamma(1|X^T,\zeta^T)\in[0,1]$ replacing the binary-valued detector. Note that Lemma \ref{Thm:min typeII error fixed} holds for any randomized detector as well, as shown in Supplementary~\ref{App:converse min error fixed}. 
The following result establishes the necessary and sufficient conditions on fixed discrete common randomness such that an optimal construction exists. To state the result succinctly, we define the \emph{heavy} and \emph{light} sets of atoms of $P^*_{X^T}$: $\calH \coloneqq \{x^T\in\calX:P_{X^T}^*(x^T)\ge\alpha\}$ and $\calL \coloneqq \{x^T\in\calX:P_{X^T}^*(x^T)<\alpha\}$.
\begin{theorem}[Fixed discrete common randomness: necessary and sufficient conditions for attaining the lower bound in Lemma~\ref{Thm:min typeII error fixed}]
\label{thm:randomized-discrete-optimal-construction}
Fix an LLM with generation distribution $Q_{X^T}$, $\alpha \in [0,1]$ and $\epsilon>0$. 
There exist a coupling $P_{X^T,\zeta^T}\in\Pi(P_{X^T}^*,P_{\zeta^T})$ and a randomized detector $\gamma$
that attains the lower bound in Lemma~\ref{Thm:min typeII error fixed}
if and only if there exist subsets of $\calZ^T=\mathrm{supp}(P_{\zeta^T})$ satisfying the following conditions:
\begin{enumerate}
\item Pairwise disjoint sets $E_{x^T}\subseteq\calZ^T$, ${x^T}\in\calH$, satisfying $\alpha\le P_{\zeta^T}(E_{x^T})\le P_{X^T}^*(x^T)$ for all ${x^T}\in\calH$.
\item With $\calR=\calZ^T\setminus\bigcup_{{x^T}\in\calH}E_{x^T}$, sets
$S_{x^T}\subseteq\calR$, ${x^T}\in\calL$, satisfying $P_{\zeta^T}(S_{x^T})\le\alpha$ for all $ {x^T}\in\calL$,
and the capacity inequalities $\sum_{{x^T}\in A}P_{X^T}^*(x^T)\le P_{\zeta^T}\!\left(\bigcup_{{x^T}\in A}S_{x^T}\right)$ for all $A\subseteq\calL$.
\end{enumerate}
The sets $S_{x^T}$ need not be disjoint. When these conditions hold, a suitable
coupling can be constructed so that the detector
\begin{equation}
\gamma(1\mid x^T,\zeta^T)=
\begin{cases}
\displaystyle\frac{\alpha}{P_{\zeta^T}(E_{x^T})}\mathbbm{1}\{{\zeta^T}\in E_{x^T}\},&{x^T}\in\calH,\\[6pt]
\mathbbm{1}\{{\zeta^T}\in S_{x^T}\},&{x^T}\in\calL
\end{cases}
\label{eq:detector}
\end{equation}
attains the lower bound in Lemma \ref{Thm:min typeII error fixed}.
\end{theorem}


The proof of Theorem~\ref{thm:randomized-discrete-optimal-construction} is deferred to Supplementary~\ref{App: pf of thm randomized-opt-cons}. Theorem~\ref{thm:randomized-discrete-optimal-construction} presents the collection of all possible discrete $P_{\zeta^T}$ that permits the existence of jointly optimal embedder-detector construction. A simple subset of it provides a sufficient condition: if $\calZ^T$ has a partition $\{B_{x^T}:{x^T}\in\calX\}$ such that $P_{\zeta^T}(B_{x^T})=P_{X^T}^*(x^T)$, then the lower bound is attained by $P_{X^T,\zeta^T}(x^T,\zeta^T)=P_{\zeta^T}(\zeta^T)\mathbbm{1}\{{\zeta^T}\in B_{x^T}\}$ and $\gamma(1 | x^T,\zeta^T)=\frac{m(x^T)}{P_{X^T}^*(x^T)}\mathbbm{1}\{{\zeta^T}\in B_{x^T}\}$.
A toy example of the achievability is illustrated in the column (b) of Figure \ref{fig:fixed CR example}. Consider the same setting as the toy example for Theorem \ref{thm:continuous-optimal-construction} but with a fixed, discrete uniform common randomness $P_\zeta=(\frac{1}{3},\frac{1}{3},\frac{1}{6},\frac{1}{6})$. The optimal coupling is constructed by setting $P_{X,\zeta}^*(a,1)=\frac{1}{3}$, $P_{X,\zeta}^*(a,4)=\frac{1}{6}$, $P_{X,\zeta}^*(b,2)=\frac{1}{3}$, and $P_{X,\zeta}^*(c,3)=\frac{1}{6}$, with all other entries equal to zero. This construction achieves a worst-case Type-I error of $\frac{1}{4}$ and a Type-II error of $\frac{1}{3}$, attaining the lower bound in Lemma~\ref{Thm:min typeII error fixed}. Several remarks are as follows.

\textbf{(i) Randomization relaxes atomic constraints.} Compared to the deterministic setting, the key advantage of randomized detectors is that they relax the restriction imposed by the atomic structure of discrete $P_{\zeta^T}$, but not every $P_{\zeta^T}$ permits the existence of joint optimal construction. Specifically, the atoms must admit mutually disjoint sets $E_{x^T}$ for the high-probability text sequences, each with mass between $\alpha$ and $P_{X^T}^*(x^T)$. The remaining atoms must accommodate the low-probability text sequences within support sets $S_{x^T}$ of mass at most $\alpha$, with sufficient capacity for every collection of such sequences. Randomization remains essential in general: on an exclusive set $E_{x^T}$ with mass greater than $\alpha$, the constructed detector accepts with probability $\frac{\alpha}{P_{\zeta^T}(E_{x^T})}$, thereby matching the false-alarm budget. If each high-probability text sequence has such an exclusive subset of exact mass $\alpha$, a deterministic detector also suffices.

\textbf{(ii) Structure and model-awareness of the optimal construction.} The optimal construction reveals that: (a) the coupling assigns exclusive atom sets to high-probability text sequences, allocates low-probability sequences within their admissible support sets, and uses the remaining mass to complete the marginals; (b) the detector accepts on $E_{x^T}$ with probability $\frac{\alpha}{P_{\zeta^T}(E_{x^T})}$, accepts on $S_{x^T}$ with probability one, and rejects elsewhere, ensuring a detection contribution of $m(x^T)$ for each sequence while respecting the false-alarm constraint; and (c) both the coupling and the detector are model-aware, since the classification of sequences and the selection of these sets depend on $P_{X^T}^*$, as well as on $P_{\zeta^T}$ and $\alpha$.

\paragraph{Deterministic Detector} 
When the detector $\gamma$ is constrained to be deterministic, as is common in practice, the lower bound can still be attained if $P_{\zeta^T}$ is chosen as an appropriate uniform distribution based on the structure of $Q_{X^T}$. The following result shows that, under such a choice of uniform $P_{\zeta^T}$, an optimal construction always exists.

\begin{assumption}[Mass structure]\label{Ass: rational mass}
Assume that there exists a constant $\mu \in (0,1)$ and a collection of positive integers $\{k_{x^T}\}_{x^T\in\mathcal V^T}$ such that $m(x^T)=k_{x^T}\mu$ for all $x^T\in\mathcal V^T$.
Equivalently, all $\{m(x^T)\}$ are rational and share a common denominator $\mu^{-1}$.
\end{assumption}

Under Assumption 1, let the auxiliary alphabet $\mathcal Z^T$ satisfy $|\mathcal Z^T| = \sum_{x^T\in\mathcal V^T} k_{x^T}$,
and define the uniform distribution $P_{\zeta^T}^{\mathrm{uni}}$ on $\mathcal Z^T$ as
$P_{\zeta^T}^{\mathrm{uni}}(\zeta^T)=\mu$, for all $ \zeta^T\in\mathcal Z^T$. Hence, the probability masses of $P_{\zeta^T}^{\mathrm{uni}}$ can be partitioned into disjoint subsets whose total masses exactly match $\{m(x^T)\}_{x^T\in\mathcal V^T}$.


\begin{theorem}[Discrete uniform common randomness $P_{\zeta^T}=P_{\zeta^T}^{\mathrm{uni}}$: existence of jointly optimal watermark embedder and deterministic detector]\label{thm:discrete-optimal-construction}
   Suppose that Assumption \ref{Ass: rational mass} holds. Fix an LLM with distribution $Q_{X^T}$, $\alpha \in [0,1]$ and $\epsilon>0$. 
Let $P_{\zeta^T}=P_{\zeta^T}^{\mathrm{uni}}$ on the discrete space $\mathcal Z^T$. 
Then there exist
\begin{enumerate}
    \item a deterministic detector $\gamma^* : \calV^T \times \mathcal Z^T \to \{0,1\}$, and
    \item a coupling $P_{X^T,\zeta^T}^*$ of $P^*_{X^T}$ (cf.~\eqref{Eq: opt X marginal}) and $P_{\zeta^T}^{\mathrm{uni}}$, 
\end{enumerate}
such that $\sup_{Q_{X^T}} \beta_0(\gamma^*,Q_{X^T},P_{\zeta^T}) \le \alpha$ and $\beta_1(\gamma^*,P_{X^T,\zeta^T}^*) = \sum_{x^T\in\calV^T} \bigl(P^*_{X^T}(x^T)- \alpha\bigr)_+$.
Hence, $\gamma^*$ and $P_{X^T,\zeta^T}^*$ attain the lower bound of the optimal value of \eqref{Eq: opt-fixed} in Lemma \ref{Thm:min typeII error fixed}.
\end{theorem}

A toy example of the achievability is illustrated in the column (c) of Figure \ref{fig:fixed CR example}. Consider the same toy example setting but with a special uniform common randomness $P_{\zeta}^{\mathrm{uni}}=\Unif([1:12])$. The resulting quantities in \eqref{Eq: m(x)} are $m(\text{a})=\frac{1}{4},m(\text{b})=\frac{1}{4},m(\text{c})=\frac{1}{6}$. Therefore, the common denominator is $12$ and $P_{\zeta}^{\mathrm{uni}}(\zeta)=\frac{1}{12}$ for all $\zeta$. As shown in the table in column (c) of Figure \ref{fig:fixed CR example}, the optimal coupling $P^*_{X,\zeta}$ is constructed by moving $m(\text{a})$ amount of mass from $P_X^*(\text{a})$ to $\{P_{\zeta}^{\mathrm{uni}}(\zeta)\}_{\zeta\in[1:3]}$, $m(\text{b})$ amount of mass from $P_X^*(\text{b})$ to $\{P_{\zeta}^{\mathrm{uni}}(\zeta)\}_{\zeta\in[4:6]}$, all mass from $P_X^*(\text{c})$ to $\{P_{\zeta}^{\mathrm{uni}}(\zeta)\}_{\zeta\in\{7,8\}}$, and the rest of mass from $P^*_X$ to $\{P_{\zeta}^{\mathrm{uni}}(\zeta)\}_{\zeta\in[9:12]}$.
 The corresponding acceptance region for $\rmH_1$ is $\{(\text{a},\zeta)\}_{\zeta\in[1:3]}\cup\{(\text{b},\zeta)\}_{\zeta\in[4:6]}\cup\{(\text{c},\zeta)\}_{\zeta\in\{7,8\}}$, shadowed by blue in Figure \ref{fig:fixed CR example}. The proof Theorem \ref{thm:discrete-optimal-construction} is deferred to Appendix \ref{App: pf of thm discrete-opt-cons}. In that appendix, we also show that for general discrete distributions $P_{\zeta^T}$, it may not always be feasible to construct an optimal scheme that achieves the lower bound in Lemma \ref{Thm:min typeII error fixed}. In this case, the optimization problem \eqref{Eq: opt-fixed} reduces to a nontrivial combinatorial problem. However, for the special case of uniform common randomness, the construction recovers the continuous $P_{\zeta^T}$ setting.

The key insights from the optimal construction are as follows:
(i) the optimal deterministic detector $\gamma^*$ is dependent on the original LLM distribution $Q_{X^T}$ through $\{m(x^T)\}$, and maps each token sequence $x^T$ to a subset of $\mathcal Z^T$, mirroring the structure in Theorem \ref{thm:continuous-optimal-construction};
(ii) the jointly optimal coupling $P_{X^T,\zeta^T}^*$ exhibits the same structural form as in Theorem \ref{thm:continuous-optimal-construction}.

\subsection{Discussions}
Overall, the above results reveal several inherent limitations of the fixed common randomness setting. First, the optimal detector $\gamma^*$ depends explicitly on both the original LLM distribution $Q_{X^T}$ (through $P_{X^T}^*$) and the prescribed common randomness $P_{\zeta^T}$, and in general may not admit a simple closed-form expression. Second, when the detector is restricted to be deterministic, the lower bound in Lemma \ref{Thm:min typeII error fixed} may fail to be attainable for arbitrary discrete $P_{\zeta^T}$ due to the indivisibility of atoms.



Despite these limitations, the lower bound in Lemma~\ref{Thm:min typeII error fixed} is tight for fixed continuous common randomness and for fixed discrete distributions satisfying the conditions in Theorem~\ref{thm:randomized-discrete-optimal-construction}. In these cases, there exist jointly optimal constructions of watermark embedders and detectors (possibly randomized) that attain this bound. This confirms that the lower bound characterizes the fundamental limit of the minimum Type-II error in these settings.
The remaining gap is therefore constructive. In particular, translating the existence results into explicit and practically implementable embedder--detector pairs may be challenging, and the optimal constructions above generally require model-aware detectors.

These observations motivate a more general formulation in which the common randomness $P_{\zeta^T}$ is no longer fixed but can be optimized. This introduces an additional degree of freedom specific to LLM watermarking, enabling the joint design of the watermark embedding, detector, and common randomness. Such flexibility is typically absent in classical coding problems, where the underlying randomness is fixed by the channel or code construction.

\section{Optimal Watermarking Scheme with Optimizable Common Randomness}
In this section, we move on to the general setting where the common randomness can be further optimized, and aim to solve the optimization in \eqref{Eq: opt-optimizable}. The detector is restricted to be deterministic, as specified in the problem formulation.  We first show that the lower bound for the minimum Type-II error in \eqref{Eq: opt-optimizable} is the same as that of \eqref{Eq: opt-fixed}. We then identify the jointly optimal watermark embedder and detector that attains this lower bound, indicating that it represents the universally minimum Type-II error under optimizable common randomness. This is section is partly based on results presented in the conference version \cite{he2025theoretically}.
\begin{theorem}[Universally Minimum Type-II Error under Optimizable Common Randomness]\label{Thm:min type-II error opt}
     Given any LLM with generation distribution $Q_{X^T}$, $\alpha\in[0,1]$ and $\epsilon\geq 0$,  the universally minimum Type-II error attained from \eqref{Eq: opt-optimizable} is equal to the lower bound in Lemma \ref{Thm:min typeII error fixed}: $$\beta_{1,\opt}^*(Q_{X^T}, \alpha, \epsilon)=\min_{\substack{P_{X^T}:  \sD(P_{X^T}, Q_{X^T} )\leq \epsilon}}\sum\nolimits_{x^T}(P_{X^T}(x^T)- \alpha)_+.$$
\end{theorem}

The converse proof of Theorem \ref{Thm:min type-II error opt} is identical to that in Supplementary~\ref{App:converse min error fixed} and applies to both discrete and continuous common randomness $P_{\zeta^T}$. We next present an achievability construction of the watermark embedder and detector, thereby completing the proof of the universally minimum Type-II error.

Theorem \ref{Thm:min type-II error opt} shows that the lower bound in Lemma \ref{Thm:min typeII error fixed} characterizes not only the fundamental detection limit under fixed common randomness, but also under optimizable common randomness. 

Moreover, we observe that $\beta_{1,\opt}^*(Q_{X^T},\alpha,\epsilon)$ coincides with the minimum Type-II error reported in \cite[Theorem 3.2]{huang2023towards}, which was shown to be optimal for their specific detector. In contrast, our results establish that this quantity is the universally minimum Type-II error over all possible watermark embedders and detectors when the common randomness is optimized, implying that their detector is a member of the class of optimal detectors characterized below.

\begin{theorem}[Optimizable common randomness: jointly optimal watermarking embedder and detector] \label{Thm: optimal detector}
 The class of optimal detectors is given by
 \begin{align}
        \!\!\!\gamma^*\in \Gamma^* \!\coloneqq \! \big\{& \gamma \mid \gamma(X^T,\zeta^T)= \mathbbm{1}\{X^T=g(\zeta^T)\}, \nn\\
        &\!\!\!\text{ for some surjective } g:\calZ^T\to  \cV^T\cup \{s_0\} \big\},
    \label{Eq: opt detector class}
\end{align}
where $\calZ^T=g^{-1}(\calV^T)\cup \{\tilde{\zeta}^T\}$ and $\tilde{\zeta}^T$ is the unique preimage of $s_0$. Here, \(s_0\) denotes a rejection symbol, and \(\tilde{\zeta}^T\) is a redundant auxiliary sequence.

Given any $(Q_{X^T},\alpha,\epsilon)$ and a detector $\gamma^* \in\Gamma^*$, the jointly optimal watermark embedder output coupling is given as follows:   for any $x^T\in \calV^T, P_{X^T,\zeta^T}^*(x^T, \zeta^T)$ satisfies $\sum_{\zeta^T}P_{X^T,\zeta^T}^*(x^T, \zeta^T)=P^*_{X^T}(x^T)$ and
\begin{equation}\label{Eq: opt cond distr}
    \begin{cases}
         \sum_{\zeta^T}   P_{X^T,\zeta^T}^*(x^T,\zeta^T) \gamma(x^T,\zeta^T) =m(x^T), & \\
          P_{X^T,\zeta^T}^*(x^T,\zeta^T) =\big(P_{X^T}^*(x^T)- \alpha \big)_+, & \text{if }  \zeta^T=\tilde{\zeta^T};\\
          P_{X^T,\zeta^T}^*(x^T,\zeta^T)=0, & \text{otherwise.}
    \end{cases}
\end{equation}
Hence, the constructed pair $(\gamma^*, P_{X^T,\zeta^T}^*)$ achieves $\sup_{Q_{X^T}} \beta_0(\gamma^*,Q_{X^T},P^*_{\zeta^T}) \le \alpha$ and attains the universally minimum Type-II error $\beta_1(\gamma^*,P_{X^T,\zeta^T}^*) =\beta_{1,\opt}^*(Q_{X^T},\alpha,\epsilon)$. 
\end{theorem}

The proof of Theorem \ref{Thm: optimal detector} is deferred to Supplementary~\ref{App: pf of Thm: optimal detector}. We highlight several remarks to further interpret the optimal construction.

\textbf{(i) Illustration of the surjective mapping.}
To provide intuition for the surjective function $g$ in the optimal detector, consider a simple example with $\mathcal V^T=\{a,b,c\}$. Define an extended set $\mathcal S=\mathcal V^T \cup \{\#\}=\{a,b,c,\#\}$ and construct a surjective mapping $g:\mathcal Z^T\to\mathcal S$ from an auxiliary set $\mathcal Z^T=\{1,2,3,4,5,6\}$. For instance, let $g(1)=a$, $g(2)=g(3)=b$, $g(4)=g(5)=c$, and $g(6)=\#$. Here, the symbol $\#$ corresponds to the rejection symbol.

\textbf{(ii) Structure of the optimal common randomness.}
We observe that it suffices to construct a discrete optimal common randomness $P_{\zeta^T}^*$, which reduces computational complexity in practice. The induced $P_{\zeta^T}^*$ can be any distribution on $\mathcal Z^T$ satisfying
\begin{align}
\sum_{\zeta^T:\,\gamma(x^T,\zeta^T)=1} P_{\zeta^T}^*(\zeta^T)
&=
P_{X^T}^*(x^T)\wedge \alpha, \quad \forall x^T\in\mathcal V^T,
\\
P_{\zeta^T}^*(\tilde{\zeta}^T)
&=
\sum_{x^T\in\mathcal V^T}\bigl(P_{X^T}^*(x^T)-\alpha\bigr)_+.
\end{align}

\textbf{(iii) Adaptivity and advantage over fixed common randomness.}
\begin{figure}[t]
    \centering
    \includegraphics[trim={0 .3cm .6cm .4cm},  width=.8\linewidth]{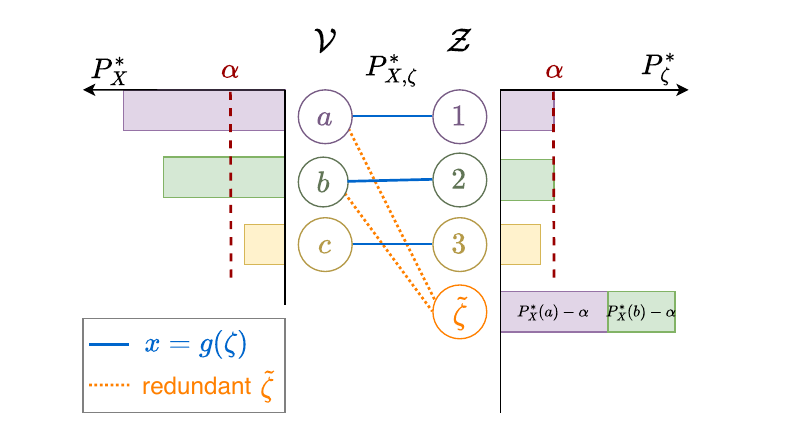} \caption{A toy example of the optimal detector and coupling when $T=1$. Links between $\calV$ and $\calZ$ suggest $P_{X,\zeta}^*> 0$. }
    \label{Fig: watermarking}
    \vspace{-10pt}
\end{figure}
To better understand the optimal coupling, we provide a detailed illustration in Figure \ref{Fig: watermarking}. A key observation is that both the optimal watermark-induced distributions $P_{X^T|\zeta^T}^*$ and the optimal common randomness $P_{\zeta^T}^*$ are inherently \emph{adaptive} to the original LLM output distribution $Q_{X^T}$. This suggests that, to maximize detection performance with a deterministic and model-agnostic detector, the distribution of common randomness should be generated tailored to $Q_{X^T}$. 

This stands in contrast to existing watermarking schemes, which typically sample the auxiliary sequence from a fixed uniform distribution, independent of the LLM NTP distribution. 
Moreover, compared to the fixed common randomness setting, optimizing $P_{\zeta^T}$ eliminates the dependence of the optimal detector $\gamma^*$ on $Q_{X^T}$ by allowing a prescribed mapping $g$: with the common random sequence $\zeta^T$ shared directly through a secret key, the detector applies a fixed deterministic rule based only on $(X^T,\zeta^T)$, while still attaining the lower bound on the minimum Type-II error in Lemma \ref{Thm:min typeII error fixed}.

\textbf{(iv) Role of the redundant auxiliary sequence.}
Consistent with the fixed common randomness construction, controlling the worst-case false positive rate requires introducing a \emph{redundant} auxiliary sequence $\tilde{\zeta}^T \in \mathcal Z^T$ such that $\gamma^*(x^T,\tilde{\zeta}^T)=0$ for all $x^T$. This sequence plays the same role as the redundant bucket $B_0$, absorbing the excess mass of $P_{X^T}^*$ above $\alpha$. In particular, if $P_{X^T}^*(x^T)>\alpha$ (corresponding to low-entropy text), part of its mass must be mapped to $\tilde{\zeta}^T$, making such sequences harder to detect as watermarked. Consequently, the optimal scheme $P_{X^T,\zeta^T}^*$ is particularly effective at controlling the false positive rate for low-entropy texts.

\section{Optimal Watermark-Induced Coupling with Fixed Uniform Common Randomness and a Deterministic Model-Agnostic Detector}~\label{Sec: opt watermark when uniform}
In this section, we consider a practical setting in which the common randomness is fixed to be uniform and the detector is fixed and deterministic, independent of the original LLM distribution $Q_{X^T}$. This setting reflects the design of many existing watermarking schemes, where the auxiliary randomness and detector are prescribed in advance rather than jointly optimized with the watermark embedder. Our goal is to highlight the performance gap incurred by this lack of joint optimization, and thereby illustrate the potential benefit of adaptive, theoretically optimal watermark design.

Specifically, we consider a uniform common randomness $P_{\zeta^T}=\Unif(\mathcal Z^T)$ on $\mathcal Z^T \supseteq \mathcal V^T$, and fix the detector as $\gamma(X^T,\zeta^T)=\mathbbm{1}\{X^T=\zeta^T\}$,
which belongs to the class $\Gamma^*$ given in Theorem \ref{Thm: optimal detector} with the associated mapping $g$ being the identity. Under this setting, we aim to solve the following optimization problem, corresponding to \eqref{Eq: opt-fixed} with a fixed detector:
\begin{align}
&\beta_{1,\mathrm{exp}}^*(\gamma,Q_{X^T},P_{\zeta^T},\alpha,\epsilon)=\inf_{P_{X^T|\zeta^T}} \beta_1(\gamma,P_{X^T,\zeta^T}) \tag{P3} \label{Eq: opt fix gamma}\\
&\text{s.t.} \quad
\sup_{Q_{X^T}} \beta_0(\gamma,Q_{X^T},P_{\zeta^T}) \le \alpha,
\;\;
\sD(P_{X^T}, Q_{X^T}) \le \epsilon. \nn
\end{align}

The following results characterize the optimal watermark-induced coupling on $\mathcal V^T \times \mathcal Z^T$ that solves \eqref{Eq: opt fix gamma} under the fixed uniform common randomness and deterministic detector. This construction can be extended to other valid surjective mappings $g$ defined in Theorem \ref{Thm: optimal detector}.

\begin{lemma}[Optimal watermark-induced coupling under uniform $P_{\zeta^T}$ and $\gamma=\mathbbm{1}\{X^T=\zeta^T\}$]\label{Lem: opt watermark when uniform}
    Given a uniform common randomness $P_{\zeta^T}=\Unif(\calZ^T)$ on $\calZ^T\supseteq\calV^T$, a detector $\gamma=\mathbbm{1}\{X^T=\zeta^T\}$,  $1\geq \alpha\geq \frac{1}{|\calZ|^T}$ and $\epsilon\geq 0$, the minimum Type-II error attained from \eqref{Eq: opt fix gamma} is 
    \begin{align}
    &\beta_{1,\text{exp}}^*(\gamma,Q_{X^T},P_{\zeta^T},\alpha,\epsilon) \nn\\
    &=\min_{\substack{P_{X^T}:\rvD(P_{X^T} , Q_{X^T} )\leq \epsilon
    }} \sum_{x^T}\bigg(P_{X^T}(x^T)-\frac{1}{|\calZ^T|} \bigg)_+,
    \end{align}
 where the minimum of the left-hand side is achieved by the watermarked text distribution
\begin{equation}
    P^\dagger_{X^T}\coloneqq\argmin_{\substack{P_{X^T}:\rvD(P_{X^T} , Q_{X^T} )\leq \epsilon
    }} \sum_{x^T}\bigg(P_{X^T}(x^T)-\frac{1}{|\calZ^T|} \bigg)_+.
\end{equation}
 The optimal watermark-induced coupling $P^*_{X^T,\zeta^T}(x^T,\zeta^T)\in \Pi(P^\dagger_{X^T},P_{\zeta^T})$ that achieves this minimum Type-II error of \eqref{Eq: opt fix gamma} is
    \begin{equation}
    \begin{aligned}
        &P^*_{X^T,\zeta^T}(x^T,\zeta^T) \nn\\
        &=\begin{cases}
        P^\dagger_{X^T}(x^T) \wedge \frac{1}{|\calZ^T|}, &\text{if}\; x^T=\zeta^T;\\
        \frac{\big(P^\dagger_{X^T}(x^T)-\frac{1}{|\calZ^T|}\big)_+ \cdot \big(\frac{1}{|\calZ^T|}-P^\dagger_{X^T}(\zeta^T)\big)_+}{\rvD_\TV(P^\dagger_{X^T},\Unif(\calZ^T))}, &\text{otherwise}.
    \end{cases}
    \end{aligned}
    \end{equation}

\end{lemma}

The proof of Lemma \ref{Lem: opt watermark when uniform} is deferred to Supplementary~\ref{App: pf of Lem: opt watermark when uniform}, and follows from the fact that
\[\rvD_{\TV}(\mu,\nu)=\inf_{\pi\in\Pi(\mu,\nu)} \pi(X\ne Y),\]
where $X\sim\mu$, $Y\sim\nu$, and $\Pi(\mu,\nu)$ denotes the set of all couplings of the Borel probability measures $\mu$ and $\nu$.
Note that when $P_{\zeta^T}=\Unif(\mathcal Z^T)$, if $\alpha < \frac{1}{|\mathcal Z^T|}$, the feasible region of \eqref{Eq: opt fix gamma} is empty. Therefore, the results hold for $1\geq \alpha\geq \frac{1}{|\calZ|^T}$.

Moreover, the resulting minimum Type-II error $\beta_{1,\mathrm{exp}}^*(\gamma,Q_{X^T},P_{\zeta^T},\alpha,\epsilon)$ is no smaller than the optimal value $\beta_{1,\mathrm{opt}}^*(Q_{X^T},\alpha,\epsilon)$ in Theorem \ref{Thm:min type-II error opt} and the fundamental limit in Lemma \ref{Thm:min typeII error fixed} (i.e., the minimum Type-II error achieved by Theorem \ref{thm:continuous-optimal-construction}). 
This is because, in Lemma \ref{Lem: opt watermark when uniform}, both the common randomness 
$P_{\zeta^T}$ and the detector $\gamma$ are fixed, and only the watermark embedder 
$P_{X^T|\zeta^T}$ is optimized. In contrast, Theorem \ref{thm:continuous-optimal-construction} jointly optimizes the embedder and detector for fixed common randomness, while Theorem \ref{Thm:min type-II error opt} further optimizes over the common randomness. Moreover, the fixed detector $\gamma(X^T,\zeta^T)=\mathbbm{1}\{X^T=\zeta^T\}$ differs from the optimal detector in Theorem 2, $\gamma^*(X^T,\zeta^T)=\mathbbm{1}\{\zeta^T \in B(X^T)\}$. Therefore, the gap remains even if the alphabet size $|\calZ^T|$ tends to infinity. 
In the special case where $|\mathcal Z^T|=\frac{1}{\alpha}$, the constructed watermark embedder, together with the given detector, achieves the universally minimum Type-II error:
\[
\beta_{1,\mathrm{exp}}^*(\gamma,Q_{X^T},P_{\zeta^T},\alpha,\epsilon)
=
\beta_{1,\mathrm{opt}}^*(Q_{X^T},\alpha,\epsilon).
\]
In practice, however, this condition is rarely satisfied, since $\frac{1}{|\mathcal Z|^T}=\mathcal O(10^{-4T})$ is typically much smaller than practical values of $\alpha$. This gap highlights the performance loss incurred by fixing the common randomness and detector, rather than jointly optimizing the watermark embedder, detector, and common randomness. 
Nevertheless, when $T=1$, it is possible to match the Type-I error constraint by choosing the auxiliary alphabet size $|\mathcal Z|=\frac{1}{\alpha}$, thereby mitigating the performance loss.

\section{Conclusion and Future Work}

In this paper, we developed a unified information-theoretic framework for statistical watermarking of LLMs, by casting the problem as a hypothesis testing task between generated text and common randomness. Within this framework, we jointly design the watermark embedder and detector under constraints on worst-case false positive rate and distributional distortion. 

Our main results characterize the fundamental limits of watermark detection under both fixed and optimizable common randomness. We show that the minimum achievable Type-II error is governed by a simple expression depending on the original LLM distribution, and that this limit is identical across continuous, discrete, and optimizable common randomness settings. Despite this equivalence in fundamental limits, we demonstrate that optimizable common randomness offers significant advantages in terms of structural simplicity, enabling the use of fixed deterministic detectors that attain optimal performance. 

In addition, we analyze a practical setting with fixed uniform common randomness and a model-agnostic deterministic detector, which reflects the design of many existing watermarking schemes. Our results quantify the resulting performance gap relative to the optimal construction, thereby highlighting the importance of joint optimization of the watermark embedder, detector, and common randomness.

Our work opens several promising directions for future research.
First, while our results characterize the fundamental limits and optimal structures of watermarking schemes, translating these insights into practical and scalable algorithms remains an important direction. In particular, designing efficient watermark embedding and detection algorithms that approximate the optimal coupling and achieve near-optimal performance in high-dimensional settings is of significant interest.
Second, our current analysis is conducted at the distributional level over full sequences. Extending the framework to explicitly account for the autoregressive structure of LLMs—where tokens are generated sequentially according to conditional distributions—would provide a more faithful modeling of practical generation processes. Such an extension may also enable the development of online or token-level watermarking schemes with stronger theoretical guarantees.
Third, our formulation assumes access to the underlying LLM generation distribution $Q_{X^T}$. In practice, this distribution is often unknown and only accessible through sampling or approximate models. Developing robust watermarking schemes under model uncertainty or mismatch is, therefore, an important open problem.
Finally, extending the current framework to adversarial settings, such as paraphrasing or semantic-preserving transformations, would further enhance the robustness and applicability of watermarking methods in real-world scenarios.
  
\section*{Acknowledgement}
We sincerely thank the reviewers and action editor for their meticulous reading and insightful comments that led to a significantly improved version of the paper.
This research/project is supported by the
National Natural Science Foundation of China (No.~62506265),
Guangdong Basic and Applied Basic Research Foundation (No.~2025A1515110273), 
Guangdong Provincial Key Lab of Integrated Communication, Sensing and Computation for Ubiquitous Internet of Things (No.~2023B1212010007).


\appendix

\subsection{Proof of Theorem \ref{thm:continuous-optimal-construction}}\label{App: pf of thm:cont-opt-construction}
For notational simplicity, we suppress subscripts when no confusion arises. In particular, we write $P^*(x^T)$, $P(\zeta^T)$, and $P(x^T,\zeta^T)$ for $P^*_{X^T}(x^T)$, $P_{\zeta^T}(\zeta^T)$, and $P_{X^T,\zeta^T}(x^T,\zeta^T)$, respectively.

Recall $m(x^T)\coloneqq P^*(x^T)\wedge \alpha \le P^*(x^T), \qquad \forall x^T$,
and we have
\[
\sum_{x^T\in\calV^T} m(x^T) \le \sum_{x^T\in\calV^T} P^*(x^T)=1.
\]
Because $P_{\zeta^T}$ is nonatomic, there exists a collection of pairwise disjoint measurable sets
$\{B(x^T)\}_{x^T\in\calV^T}$ such that
\[
P_{\zeta^T}(B(x^T)) = m(x^T), \qquad \forall x^T \in \calV^T.
\]
Let $B_0 := \mathcal Z^T \setminus \bigcup_{x^T\in\calV^T} B(x^T)$
so that $P_{\zeta^T}(B_0)=1-\sum_{x^T} m(x^T)$.

We now define the deterministic detector
\[
\gamma^*(x^T,\zeta^T)
=
\mathbbm{1}\{\zeta^T \in B(x^T)\},
\]
that is, the detector declares $\rmH_1$ if and only if $\zeta^T \in B(x^T)$.

We next construct a coupling $P^*(x^T,\zeta^T)$ as follows. Define
\[
r(x^T) := P^*(x^T)-m(x^T), \qquad \forall x^T.
\]
Then set: for any measurable set $A \subseteq \calZ^T$,
\begin{align}
        &P^*(x^T,A)\\
        &=\begin{cases}P_{\zeta^T}(A\cap B(x^T))\\
       \qquad + \dfrac{r(x^T)}{P_{\zeta^T}(B_0)}\,P_{\zeta^T}(A\cap B_0), & P_{\zeta^T}(B_0)>0,\\
        P_{\zeta^T}(A\cap B(x^T)), & P_{\zeta^T}(B_0)=0.
        \end{cases}
\end{align}
If $P_{\zeta^T}(B_0)=0$, then necessarily $r(x^T)=0$ for all $x^T$, and the second term in the first line is absent.

In Supplementary \ref{Supp: supp to pf of thm:cont-opt-construction}, We verify that $P^*$ is indeed a coupling of $P^*_{X^T}$ and $P_{\zeta^T}$.




\paragraph{Worst-case Type-I error.}
For any distribution $Q_{X^T}$,
\begin{align}
\beta_0(\gamma^*,Q_{X^T},P_{\zeta^T})
&=
\sum_{x^T} Q_{X^T}(x^T)\,P_{\zeta^T}(B(x^T))\\
&=
\sum_{x^T} Q_{X^T}(x^T)\,m(x^T).
\end{align}
Since $m(x^T)\le \alpha$ for all $x^T$,
\begin{align}
&\beta_0(\gamma^*,Q_{X^T},P_{\zeta^T})
\le
\sum_{x^T} Q_{X^T}(x^T)\alpha
=
\alpha \nn\\
&\Longrightarrow \quad \sup_{Q_{X^T}} \beta_0(\gamma^*,Q_{X^T},P_{\zeta^T}) \le \alpha.
\end{align}

\paragraph{Type-II error.}
Under the coupling $P^*_{X^T,\zeta^T}$,
\begin{align}
1-\beta_1(\gamma^*,P^*_{X^T,\zeta^T})
&=
\sum_{x^T}\int_{\{\zeta^T:\,\gamma^*(x^T,\zeta^T)=1\}} P^*(x^T,\zeta^T) ~d\zeta^T \nn\\
&=
\sum_{x^T} P^*\bigl(x^T,B(x^T)\bigr).
\end{align}
By construction,
\[
P^*\bigl(x^T,B(x^T)\bigr)=P_{\zeta^T}(B(x^T))=m(x^T),
\]
and thus
\[
1-\beta_1(\gamma^*,P^*_{X^T,\zeta^T})
=
\sum_{x^T} m(x^T)
=
\sum_{x^T}\bigl(P_{X^T}(x^T)\wedge \alpha\bigr).
\]
Equivalently,
\[
\beta_1(\gamma^*,P^*_{X^T,\zeta^T})
=
1-\sum_{x^T}\bigl(P^*_{X^T}(x^T)\wedge \alpha\bigr)=\sum_{x^T}\bigl(P^*_{X^T}(x^T)-\alpha\bigr)_+.
\]

Finally, this matches the lower bound in Lemma \ref{Thm:min typeII error fixed} established previously for the fixed-$P_{\zeta^T}$ problem, so the construction is optimal.

\subsection{Proof of Theorem \ref{thm:discrete-optimal-construction}}\label{App: pf of thm discrete-opt-cons}

Similar to Appendix~\ref{App: pf of thm:cont-opt-construction}, we suppress subscripts when no confusion arises. 
By Assumption~\ref{Ass: rational mass},
\[
m(x^T)\coloneqq P^*(x^T)\wedge \alpha=k_{x^T}\mu, \qquad \forall x^T,
\]

In the optimization problem \eqref{Eq: opt-fixed}, we notice that if and only if $\sum_{\zeta^T}P(x^T,\zeta^T)\gamma(x^T,\zeta^T)=m(x^T)$ for all $x^T$, the lower bound in Lemma \ref{Thm:min typeII error fixed} can be achieved.

We aim to construct a deterministic detector $\gamma$ defined via an acceptance region $\calA_1\subseteq \calV^T\times \calZ^T$, i.e., $\gamma(x^T,\zeta^T)=\mathbbm{1}\{(x^T,\zeta^T)\in\calA_1\}$. Let $B(x^T)=\{\zeta^T: (x^T,\zeta^T)\in\calA_1\}$ for all $x^T\in\calV^T$, where $\{B(x^T)\}_{x^T\in\calV^T}$ are not necessarily disjoint.
Based on the deterministic detector $\gamma$, we aim to construct a coupling $P(x^T,\zeta^T)\in\Pi(P^*_{X^T},P_{\zeta^T})$ such that 
\begin{equation}
\sum_{\zeta^T\in B(x^T)}P(x^T,\zeta^T)=m(x^T), \qquad \forall x^T. \label{Eq:opt coupling constraint}
\end{equation}

\vspace{0.5em}
\noindent
\textbf{Step 1: Construction of sets $B(x^T)$ and deterministic detector $\gamma^*$.}
The coupling constraint \eqref{Eq:opt coupling constraint} implies that the marginal $P_{\zeta^T}(B(x^T))$ should have enough volume to accept $m(x^T)$ mass. 
We first construct (not necessarily disjoint) sets $B(x^T) \subseteq \mathcal{Z}^T$ such that
\[
\alpha \geq P_{\zeta^T}(B(x^T)) \ge m(x^T), \quad \forall x^T,
\]
where the first inequality follows from the worst-case Type-I error constraint.
In particular, one may choose $P_{\zeta^T}(B(x^T)) = m(x^T)$ whenever possible.

Such a construction always exists in the continuous case by selecting measurable subsets of $\mathcal{Z}^T$ with prescribed probability mass. In the discrete case, feasibility depends on the structure of $P_{\zeta^T}$ and $\{m(x^T)\}$.


\paragraph{Uniform $P_{\zeta^T}(\zeta^T)=\mu$ for all $\zeta^T$} There always exists a collection of pairwise disjoint sets 
$\{B(x^T)\}_{x^T\in\calV^T}$ such that
\[
P_{\zeta^T}(B(x^T))=|B(x^T)|\mu = m(x^T), \qquad \forall x^T \in \calV^T.
\]
Let
\[
B_0 := \mathcal Z^T \setminus \bigcup_{x^T\in\calV^T} B(x^T),
\]
so that
\[
P_{\zeta^T}(B_0)=1-\sum_{x^T} m(x^T).
\]
We then set $\gamma ^*(X^T,\zeta^T)=\mathbbm{1}\{\zeta^T\in B(X^T)\}$.

\paragraph{Any other discrete $P_{\zeta^T}$} In any other general discrete distribution $P_{\zeta^T}$, we may not be able to find disjoint sets $\{B(x^T)\}$ such that $P_{\zeta^T}(B(x^T)) = m(x^T)$ for all $x^T$.  There are two possible ways to construct $\{B(x^T)\}$ but they may not always be feasible:
\begin{enumerate}
    \item Let $B(x^T)=B$ for all $x^T$ such that $\alpha \geq P_{\zeta^T}(B) \geq \max_{x^T} m(x^T)$;
    \item Pick overlapping $\{B(x^T)\}$ such that $\alpha \geq P_{\zeta^T}(B(x^T)) \ge m(x^T)$ for all $x^T$.
\end{enumerate}

\vspace{0.5em}
\noindent
\textbf{Step 2: Coupling construction.}
Based on the constructed (not necessarily disjoint) sets $\{B(x^T)\}_{x^T\in\calV^T}$,
define
\[
P^*(x^T,\zeta^T) = M_{\mathrm{acc}}(x^T,\zeta^T) + M_{\mathrm{rej}}(x^T,\zeta^T),
\]
where $M_{\mathrm{acc}}, M_{\mathrm{rej}}\in\bbR_{\geq 0}^{|\calV^T|\times |\calZ^T|}$ are non-negative matrices, and the accepted part satisfies
\[
M_{\mathrm{acc}}(x^T,\zeta^T) = 0 \quad \text{if } \zeta^T \notin B(x^T),
\]
and
\[
\sum_{\zeta^T} M_{\mathrm{acc}}(x^T,\zeta^T) = m(x^T), \quad \forall x^T,
\]
with the column constraint
\[
\sum_{x^T} M_{\mathrm{acc}}(x^T,\zeta^T) \le P(\zeta^T), \quad \forall \zeta^T.
\]
However, such a non-negative matrix $M_{\mathrm{acc}}$ may not always exist.

\vspace{0.5em}
\noindent
\textbf{Step 3: Remaining mass.}
Define the leftover row and column masses:
\[
r(x^T) := P^*(x^T) - m(x^T), \quad
c(\zeta^T) := P(\zeta^T) - \sum_{x^T} M_{\mathrm{acc}}(x^T,\zeta^T).
\]
We then define $M_{\mathrm{rej}}(x^T,\zeta^T) = \frac{r(x^T)\,c(\zeta^T)}{1 - \sum_{x^T} m(x^T)}.$

\vspace{0.5em}
\noindent
\textbf{Step 4: Validity.}
By construction,
\[
\sum_{\zeta^T} M_{\mathrm{rej}}(x^T,\zeta^T) = r(x^T), \quad
\sum_{x^T} M_{\mathrm{rej}}(x^T,\zeta^T) = c(\zeta^T),
\]
so that $P^*(x^T,\zeta^T)$ satisfies the desired marginal constraints.

For the speical uniform distribution  $P_{\zeta^T}(\zeta^T)=\mu$ for all $\zeta^T$, we can construct $P^*(x^T,\zeta^T)$ using the same way as Theorem \ref{thm:continuous-optimal-construction}.

\vspace{0.5em}
\noindent
\textbf{Step 5: Errors.} 
Following similar steps in Appendix \ref{App: pf of thm:cont-opt-construction}, we can show that the worst-case Type-I error is upper bounded by $\alpha$ and the Type-II error achieves the lower bound in Lemma \ref{Thm:min typeII error fixed}.

\vspace{0.5em}
\noindent
\textbf{Feasibility condition.}
A necessary and sufficient condition for the existence of $M_{\mathrm{acc}}$ is that the constructed sets $\{B(x^T)\}$ satisfy
\[
\sum_{x^T \in S} m(x^T) \le P_{\zeta^T}\Big( \bigcup_{x^T \in S} B(x^T) \Big),
\quad \forall S \subseteq \mathcal{X}^T,
\]
which is a Hall-type condition \cite{hall1987representatives}.

\bibliographystyle{IEEEtran}
\bibliography{IT_references,sample}

@inproceedings{long2025optimized,
  title={Optimized Couplings for Watermarking Large Language Models},
  author={Long, Carol Xuan and Tsur, Dor and Verdun, Claudio Mayrink and Hsu, Hsiang and Permuter, Haim and Calmon, Flavio P},
  booktitle={2025 IEEE International Symposium on Information Theory (ISIT)},
  pages={1--6},
  year={2025},
  organization={IEEE}
}

@article{costa1983writing,
  title={Writing on dirty paper (corresp.)},
  author={Costa, Max},
  journal={IEEE transactions on information theory},
  volume={29},
  number={3},
  pages={439--441},
  year={1983},
  publisher={IEEE}
}

@inproceedings{tsurheavywater,
  title={HeavyWater and SimplexWater: Distortion-free LLM Watermarks for Low-Entropy Distributions},
  author={Tsur, Dor and Long, Carol Xuan and Verdun, Claudio Mayrink and Vithana, Sajani and Hsu, Hsiang and Chen, Chun-Fu and Permuter, Haim H and Calmon, Flavio},
  year={2025},
  booktitle={The Thirty-ninth Annual Conference on Neural Information Processing Systems}
}

@article{he2025distributional,
  title={Distributional information embedding: A framework for multi-bit watermarking},
  author={He, Haiyun and Liu, Yepeng and Wang, Ziqiao and Mao, Yongyi and Bu, Yuheng},
  journal={arXiv preprint arXiv:2501.16558},
  year={2025}
}

@incollection{hall1987representatives,
  title={On representatives of subsets},
  author={Hall, Philip},
  booktitle={Classic Papers in Combinatorics},
  pages={58--62},
  year={1987},
  publisher={Springer}
}

@book{cover2006elements,
  author    = {Cover, Thomas M. and Thomas, Joy A.},
  title     = {Elements of Information Theory},
  edition   = {2},
  publisher = {Wiley-Interscience},
  address   = {Hoboken, NJ},
  year      = {2006},
  isbn      = {9780471241959}
}

@article{shannon1949communication,
  author  = {Shannon, Claude E.},
  title   = {Communication Theory of Secrecy Systems},
  journal = {Bell System Technical Journal},
  volume  = {28},
  number  = {4},
  pages   = {656--715},
  year    = {1949},
  doi     = {10.1002/j.1538-7305.1949.tb00928.x}
}

@article{blackwell1960capacities,
  author  = {Blackwell, David and Breiman, Leo and Thomasian, A. J.},
  title   = {The Capacities of Certain Channel Classes under Random Coding},
  journal = {Annals of Mathematical Statistics},
  volume  = {31},
  number  = {3},
  pages   = {558--567},
  year    = {1960},
  doi     = {10.1214/aoms/1177705783}
}

@article{ahlswede1978elimination,
  author  = {Ahlswede, Rudolf},
  title   = {Elimination of Correlation in Random Codes for Arbitrarily Varying Channels},
  journal = {Zeitschrift f{\"u}r Wahrscheinlichkeitstheorie und Verwandte Gebiete},
  volume  = {44},
  number  = {2},
  pages   = {159--175},
  year    = {1978},
  doi     = {10.1007/BF00533053}
}

@article{gelfand1980coding,
  author  = {Gel'fand, S. I. and Pinsker, M. S.},
  title   = {Coding for Channel with Random Parameters},
  journal = {Problems of Control and Information Theory},
  volume  = {9},
  number  = {1},
  pages   = {19--31},
  year    = {1980}
}

@inproceedings{kirchenbauer2023watermark,
  title={A watermark for large language models},
  author={Kirchenbauer, John and Geiping, Jonas and Wen, Yuxin and Katz, Jonathan and Miers, Ian and Goldstein, Tom},
  booktitle={International Conference on Machine Learning},
  pages={17061--17084},
  year={2023},
  organization={PMLR}
}

@article{li2024statistical,
  title={A statistical framework of watermarks for large language models: Pivot, detection efficiency and optimal rules},
  author={Li, Xiang and Ruan, Feng and Wang, Huiyuan and Long, Qi and Su, Weijie J},
  journal={The Annals of Statistics},
  volume={53},
  number={1},
  pages={322--351},
  year={2025},
  publisher={Institute of Mathematical Statistics}
}

@book{thomas2006elements,
author = {Cover, Thomas M. and Thomas, Joy A.},
title = {Elements of Information Theory (Wiley Series in Telecommunications and Signal Processing)},
year = {2006},
isbn = {0471241954},
publisher = {Wiley-Interscience},
address = {USA}
}

@article{liu2019roberta,
  title={Roberta: A robustly optimized bert pretraining approach},
  author={Liu, Yinhan},
  journal={arXiv preprint arXiv:1907.11692},
  year={2019}
}

@article{radford2019language,
  title={Language models are unsupervised multitask learners},
  author={Radford, Alec and Wu, Jeffrey and Child, Rewon and Luan, David and Amodei, Dario and Sutskever, Ilya and others},
  journal={OpenAI blog},
  volume={1},
  number={8},
  pages={9},
  year={2019}
}

@article{zhang2022opt,
  title={Opt: Open pre-trained transformer language models},
  author={Zhang, Susan and Roller, Stephen and Goyal, Naman and Artetxe, Mikel and Chen, Moya and Chen, Shuohui and Dewan, Christopher and Diab, Mona and Li, Xian and Lin, Xi Victoria and others},
  journal={arXiv preprint arXiv:2205.01068},
  year={2022}
}

@article{touvron2023llama,
  title={Llama: Open and efficient foundation language models},
  author={Touvron, Hugo and Lavril, Thibaut and Izacard, Gautier and Martinet, Xavier and Lachaux, Marie-Anne and Lacroix, Timoth{\'e}e and Rozi{\`e}re, Baptiste and Goyal, Naman and Hambro, Eric and Azhar, Faisal and others},
  journal={arXiv preprint arXiv:2302.13971},
  year={2023}
}

@misc{gumbel2023,
  title = {Watermarking of large language models},
  howpublished = {\url{https://simons.berkeley.edu/talks/scott-aaronson-ut-austin-openai-2023-08-17}},
  year = {2023},
  note = {Accessed: 2023-08},
  author={Aaronson, Scott}
}

@article{kuditipudi2023robust,
  title={Robust distortion-free watermarks for language models},
  author={Kuditipudi, Rohith and Thickstun, John and Hashimoto, Tatsunori and Liang, Percy},
  journal={arXiv preprint arXiv:2307.15593},
  year={2023}
}

@inproceedings{christ2024undetectable,
  title={Undetectable watermarks for language models},
  author={Christ, Miranda and Gunn, Sam and Zamir, Or},
  booktitle={The Thirty Seventh Annual Conference on Learning Theory},
  pages={1125--1139},
  year={2024},
  organization={PMLR}
}

@article{huang2023towards,
  title={Towards optimal statistical watermarking},
  author={Huang, Baihe and Zhu, Banghua and Zhu, Hanlin and Lee, Jason D and Jiao, Jiantao and Jordan, Michael I},
  journal={arXiv preprint arXiv:2312.07930},
  year={2023}
}

@inproceedings{liu2024adaptive,
  title={Adaptive Text Watermark for Large Language Models},
  author={Liu, Yepeng and Bu, Yuheng},
  booktitle={Forty-first International Conference on Machine Learning},
    year={2024}
}

@article{zhao2023provable,
  title={Provable robust watermarking for {AI}-generated text},
  author={Zhao, Xuandong and Ananth, Prabhanjan and Li, Lei and Wang, Yu-Xiang},
  journal={arXiv preprint arXiv:2306.17439},
  year={2023}
}

@inproceedings{liu2024a,
  title={A Semantic Invariant Robust Watermark for Large Language Models},
  author={Aiwei Liu and Leyi Pan and Xuming Hu and Shiao Meng and Lijie Wen},
  booktitle={The Twelfth International Conference on Learning Representations},
  year={2024},
}

@article{zhao2024permute,
  title={Permute-and-Flip: An optimally robust and watermarkable decoder for {LLMs}},
  author={Zhao, Xuandong and Li, Lei and Wang, Yu-Xiang},
  journal={arXiv preprint arXiv:2402.05864},
  year={2024}
}

@inproceedings{fernandez2023three,
  title={Three bricks to consolidate watermarks for large language models},
  author={Fernandez, Pierre and Chaffin, Antoine and Tit, Karim and Chappelier, Vivien and Furon, Teddy},
  booktitle={2023 IEEE International Workshop on Information Forensics and Security (WIFS)},
  pages={1--6},
  year={2023},
  organization={IEEE}
}

@article{boroujeny2024multi,
  title={Multi-Bit Distortion-Free Watermarking for Large Language Models},
  author={Boroujeny, Massieh Kordi and Jiang, Ya and Zeng, Kai and Mark, Brian},
  journal={arXiv preprint arXiv:2402.16578},
  year={2024}
}

@article{fu2024gumbelsoft,
  title={GumbelSoft: Diversified Language Model Watermarking via the {GumbelMax-trick}},
  author={Fu, Jiayi and Zhao, Xuandong and Yang, Ruihan and Zhang, Yuansen and Chen, Jiangjie and Xiao, Yanghua},
  journal={arXiv preprint arXiv:2402.12948},
  year={2024}
}

@article{wu2023dipmark,
  title={Dipmark: A stealthy, efficient and resilient watermark for large language models},
  author={Wu, Yihan and Hu, Zhengmian and Zhang, Hongyang and Huang, Heng},
  journal={arXiv preprint arXiv:2310.07710},
  year={2023}
}

@article{qu2024provably,
  title={Provably robust multi-bit watermarking for {AI}-generated text via error correction code},
  author={Qu, Wenjie and Yin, Dong and He, Zixin and Zou, Wei and Tao, Tianyang and Jia, Jinyuan and Zhang, Jiaheng},
  journal={arXiv preprint arXiv:2401.16820},
  year={2024}
}

@article{martinian2005authentication,
  title={Authentication with distortion criteria},
  author={Martinian, Emin and Wornell, Gregory W and Chen, Brian},
  journal={IEEE Transactions on Information Theory},
  volume={51},
  number={7},
  pages={2523--2542},
  year={2005},
  publisher={IEEE}
}

@inproceedings{moulin2000information,
  title={Information-theoretic analysis of watermarking},
  author={Moulin, Pierre and O'Sullivan, Joseph A},
  booktitle={2000 IEEE International Conference on Acoustics, Speech, and Signal Processing. Proceedings (Cat. No. 00CH37100)},
  volume={6},
  pages={3630--3633},
  year={2000},
  organization={IEEE}
}

@phdthesis{chen2000design,
  title={Design and analysis of digital watermarking, information embedding, and data hiding systems},
  author={Chen, Brian},
  year={2000},
  school={Massachusetts Institute of Technology}
}

@article{brassil1995electronic,
  title={Electronic marking and identification techniques to discourage document copying},
  author={Brassil, Jack T and Low, Steven and Maxemchuk, Nicholas F and O'Gorman, Lawrence},
  journal={IEEE Journal on Selected Areas in Communications},
  volume={13},
  number={8},
  pages={1495--1504},
  year={1995},
  publisher={IEEE}
}

@article{yoo2023robust,
  title={Robust multi-bit natural language watermarking through invariant features},
  author={Yoo, KiYoon and Ahn, Wonhyuk and Jang, Jiho and Kwak, Nojun},
  journal={arXiv preprint arXiv:2305.01904},
  year={2023}
}

@article{yang2023watermarking,
  title={Watermarking text generated by black-box language models},
  author={Yang, Xi and Chen, Kejiang and Zhang, Weiming and Liu, Chang and Qi, Yuang and Zhang, Jie and Fang, Han and Yu, Nenghai},
  journal={arXiv preprint arXiv:2305.08883},
  year={2023}
}

@article{munyer2023deeptextmark,
  title={DeepTextMark: A Deep Learning-Driven Text Watermarking Approach for Identifying Large Language Model Generated Text},
  author={Munyer, Travis and Tanvir, Abdullah and Das, Arjon and Zhong, Xin},
  journal={arXiv preprint arXiv:2305.05773},
  year={2023}
}

@inproceedings{yang2022tracing,
  title={Tracing text provenance via context-aware lexical substitution},
  author={Yang, Xi and Zhang, Jie and Chen, Kejiang and Zhang, Weiming and Ma, Zehua and Wang, Feng and Yu, Nenghai},
  booktitle={Proceedings of the AAAI Conference on Artificial Intelligence},
  volume={36},
  pages={11613--11621},
  year={2022}
}

@article{sato2023embarrassingly,
  title={Embarrassingly simple text watermarks},
  author={Sato, Ryoma and Takezawa, Yuki and Bao, Han and Niwa, Kenta and Yamada, Makoto},
  journal={arXiv preprint arXiv:2310.08920},
  year={2023}
}

@inproceedings{zhang2024remark,
  title={$\{$REMARK-LLM$\}$: A Robust and Efficient Watermarking Framework for Generative Large Language Models},
  author={Zhang, Ruisi and Hussain, Shehzeen Samarah and Neekhara, Paarth and Koushanfar, Farinaz},
  booktitle={33rd USENIX Security Symposium (USENIX Security 24)},
  pages={1813--1830},
  year={2024}
}

@inproceedings{abdelnabi2021adversarial,
  title={Adversarial watermarking transformer: Towards tracing text provenance with data hiding},
  author={Abdelnabi, Sahar and Fritz, Mario},
  booktitle={2021 IEEE Symposium on Security and Privacy (SP)},
  pages={121--140},
  year={2021},
  organization={IEEE}
}

@article{wang2023towards,
  title={Towards codable text watermarking for large language models},
  author={Wang, Lean and Yang, Wenkai and Chen, Deli and Zhou, Hao and Lin, Yankai and Meng, Fandong and Zhou, Jie and Sun, Xu},
  journal={arXiv preprint arXiv:2307.15992},
  year={2023}
}

@article{fairoze2023publicly,
  title={Publicly detectable watermarking for language models},
  author={Fairoze, Jaiden and Garg, Sanjam and Jha, Somesh and Mahloujifar, Saeed and Mahmoody, Mohammad and Wang, Mingyuan},
  journal={arXiv preprint arXiv:2310.18491},
  year={2023}
}

@inproceedings{hu2023unbiased,
  title={Unbiased Watermark for Large Language Models},
  author={Hu, Zhengmian and Chen, Lichang and Wu, Xidong and Wu, Yihan and Zhang, Hongyang and Huang, Heng},
  booktitle={The Twelfth International Conference on Learning Representations},
  year={2024}
}

@article{huo2024token,
  title={Token-Specific Watermarking with Enhanced Detectability and Semantic Coherence for Large Language Models},
  author={Huo, Mingjia and Somayajula, Sai Ashish and Liang, Youwei and Zhang, Ruisi and Koushanfar, Farinaz and Xie, Pengtao},
  journal={arXiv preprint arXiv:2402.18059},
  year={2024}
}

@article{zhang2023watermarks,
  title={Watermarks in the sand: Impossibility of strong watermarking for generative models},
  author={Zhang, Hanlin and Edelman, Benjamin L and Francati, Danilo and Venturi, Daniele and Ateniese, Giuseppe and Barak, Boaz},
  journal={arXiv preprint arXiv:2311.04378},
  year={2023}
}

@article{tu2023waterbench,
  title={Waterbench: Towards holistic evaluation of watermarks for large language models},
  author={Tu, Shangqing and Sun, Yuliang and Bai, Yushi and Yu, Jifan and Hou, Lei and Li, Juanzi},
  journal={arXiv preprint arXiv:2311.07138},
  year={2023}
}

@article{ren2023robust,
  title={A robust semantics-based watermark for large language model against paraphrasing},
  author={Ren, Jie and Xu, Han and Liu, Yiding and Cui, Yingqian and Wang, Shuaiqiang and Yin, Dawei and Tang, Jiliang},
  journal={arXiv preprint arXiv:2311.08721},
  year={2023}
}

@ARTICLE{1564436,
  author={Merhav, N. and Ordentlich, E.},
  journal={IEEE Transactions on Information Theory}, 
  title={On causal and semicausal codes for joint information embedding and source coding}, 
  year={2006},
  volume={52},
  number={1},
  pages={213-226},
  doi={10.1109/TIT.2005.860428}}

@ARTICLE{4418489,
  author={Merhav, Neri and Sabbag, Erez},
  journal={IEEE Transactions on Information Theory}, 
  title={Optimal Watermark Embedding and Detection Strategies Under Limited Detection Resources}, 
  year={2008},
  volume={54},
  number={1},
  pages={255-274},
  doi={10.1109/TIT.2007.911210}}

@inproceedings{giboulot2024watermax,
title={WaterMax: breaking the {LLM} watermark detectability-robustness-quality trade-off},
author={Eva Giboulot and Teddy Furon},
booktitle={The Thirty-eighth Annual Conference on Neural Information Processing Systems},
year={2024},
url={https://openreview.net/forum?id=HjeKHxK2VH}
}

@article{liu2024survey,
  title={A survey of text watermarking in the era of large language models},
  author={Liu, Aiwei and Pan, Leyi and Lu, Yijian and Li, Jingjing and Hu, Xuming and Zhang, Xi and Wen, Lijie and King, Irwin and Xiong, Hui and Yu, Philip},
  journal={ACM Computing Surveys},
  volume={57},
  number={2},
  pages={1--36},
  year={2024},
  publisher={ACM New York, NY}
}

@inproceedings{hou2024semstamp,
  title={SemStamp: A Semantic Watermark with Paraphrastic Robustness for Text Generation},
  author={Hou, Abe and Zhang, Jingyu and He, Tianxing and Wang, Yichen and Chuang, Yung-Sung and Wang, Hongwei and Shen, Lingfeng and Van Durme, Benjamin and Khashabi, Daniel and Tsvetkov, Yulia},
  booktitle={Proceedings of the 2024 Conference of the North American Chapter of the Association for Computational Linguistics: Human Language Technologies (Volume 1: Long Papers)},
  pages={4067--4082},
  year={2024}
}

@article{wu2025survey,
  title={A survey on {LLM}-generated text detection: Necessity, methods, and future directions},
  author={Wu, Junchao and Yang, Shu and Zhan, Runzhe and Yuan, Yulin and Chao, Lidia Sam and Wong, Derek Fai},
  journal={Computational Linguistics},
  pages={1--66},
  year={2025},
  publisher={MIT Press 255 Main Street, 9th Floor, Cambridge, Massachusetts 02142, USA~…}
}

@inproceedings{wouters2024optimizing,
  title={Optimizing Watermarks for Large Language Models},
  author={Wouters, Bram},
  booktitle={International Conference on Machine Learning},
  pages={53251--53269},
  year={2024},
  organization={PMLR}
}

@article{cai2024towards,
  title={Towards better statistical understanding of watermarking LLMs},
  author={Cai, Zhongze and Liu, Shang and Wang, Hanzhao and Zhong, Huaiyang and Li, Xiaocheng},
  journal={arXiv preprint arXiv:2403.13027},
  year={2024}
}

@article{takezawa2023necessary,
  title={Necessary and sufficient watermark for large language models},
  author={Takezawa, Yuki and Sato, Ryoma and Bao, Han and Niwa, Kenta and Yamada, Makoto},
  journal={arXiv preprint arXiv:2310.00833},
  year={2023}
}

@inproceedings{
he2025theoretically,
title={Theoretically Grounded Framework for {LLM} Watermarking: A Distribution-Adaptive Approach},
author={Haiyun He and Yepeng Liu and Ziqiao Wang and Yongyi Mao and Yuheng Bu},
booktitle={The Thirty-ninth Annual Conference on Neural Information Processing Systems},
year={2025}
}

\clearpage
\section*{Supplementary Materials}

\setcounter{subsection}{0}
\renewcommand{\thesubsection}{S.\Roman{subsection}}
\renewcommand{\thesubsectiondis}{S.\Roman{subsection}.}
\setcounter{equation}{0}
\renewcommand{\theequation}{S.\arabic{equation}}

\subsection{Proof of Lemma \ref{Thm:min typeII error fixed}: Lower bound for the Minimum Type-II error}\label{App:converse min error fixed}
We first fix a arbitrary detector $\gamma:\calV^T\times \calZ^T \to \{0,1\}$ in the following proof.  According to the Type-I error constraint, we have $\forall x^T\in \calV^T$,
\begin{align}
    \alpha&\geq \max_{Q_{X^T}}\bbE_{Q_{X^T}P_{\zeta^T}}[\gamma(X^T,\zeta^T)]\geq \bbE_{\delta_{x^T}P_{\zeta^T}}[\gamma(X^T,\zeta^T)] \nn\\
    &=\bbE_{P_{\zeta^T}}[\gamma(x^T,\zeta^T)]\\
    &=\begin{cases}
        \sum_{\zeta^T}P_{\zeta^T}(\zeta^T)\gamma(x^T,\zeta^T), &\quad \calZ \text{ is discrete;} \\
        \int P_{\zeta^T}(\zeta^T)\gamma(x^T,\zeta^T) \; d \zeta^T, &\quad \calZ \text{ is continuous;}
    \end{cases}.
\end{align}

In the following, for notational simplicity, we assume that $\calZ$ is discrete. However, the derivations hold for both discrete $\calZ$ and continuous $\calZ$.
The Type-II error is given by $1-\bbE_{P_{X^T,\zeta^T}}[\gamma(X^T,\zeta^T)]$. We have
\begin{align}\label{Eq: 1-TypeII error}
    \bbE_{P_{X^T,\zeta^T}}[\gamma(X^T,\zeta^T)]=\sum_{x^T}\underbrace{\sum_{\zeta^T}P_{X^T,\zeta^T}(x^T,\zeta^T) \gamma(x^T,\zeta^T)}_{C(x^T)}, 
\end{align}
where for all $x^T\in\calV^T$, 
\begin{equation}\label{Eq: C(x) bound}
    C(x^T)\leq P_{X^T}(x^T) ~~ \text{and} ~~ C(x^T)\leq \sum_{\zeta^T}P_{\zeta^T}(\zeta^T) \gamma(x^T,\zeta^T)\leq \alpha
\end{equation}
according to the Type-I error bound. Therefore, 
\begin{align}
    \bbE_{P_{X^T,\zeta^T}}[\gamma(X^T,\zeta^T)]&=\sum_{x^T}C(x^T) \leq \sum_{x^T}(P_{X^T}(x^T)\wedge \alpha) \nn\\
    &=1-\sum_{x^T}(P_{X^T}(x^T)- \alpha)_+ \label{Eq: 1-TypeII error ub}
\end{align}
where \eqref{Eq: 1-TypeII error ub} is maximized at 
\begin{align}\label{Eq: P_X star}
    P_{X^T}^*\coloneqq \argmin_{P_{X^T}:\sD(P_{X^T}, Q_{X^T} )\leq \epsilon}\sum_{x^T}(P_{X^T}(x^T)- \alpha)_+.
\end{align}
For any $P_{X^T}$, the Type-II error is lower bounded by
\begin{align}
     \bbE_{P_{X^T,\zeta^T}}[1-\gamma(X^T,\zeta^T)]&\geq \sum_{x^T}(P_{X^T}(x^T)- \alpha)_+.
\end{align}
By plugging $P_{X^T}^*$ into this lower bound, we obtain a Type-II lower bound that holds for all  $\gamma$ and $P_{X^T,\zeta^T}$. The equality in \eqref{Eq: 1-TypeII error ub} holds if and only if each $C(x^T)=P_{X^T}(x^T)\wedge \alpha$.

In particular, if $\gamma$ is relaxed to be a randomized detector, i.e., $\gamma:\calV^T\times\calZ^T\to \calP(\{0,1\})$, we simply replace $\gamma(X^T,\zeta^T)$ with $\gamma(1|X^T,\zeta^T)$ and the proof/result remains unchanged.

Furthermore, if we choose the distortion function to be the total variation distance $\rvD_\TV$, the lower bound can be further simplified as follows.

For any $P_{X^T}$ satisfying the distortion constraint, define
\begin{equation}
  F(P_{X^T})\coloneqq\sum_{x^T} \bigg(P_{X^T}(x^T)-\alpha\bigg)_+.
\end{equation}
Define
\begin{align*}
    E(Q_{X^T};\alpha)&\coloneqq \sum_{x^T}(Q_{X^T}(x^T)-\alpha)_+,\\
S(Q_{X^T};\alpha)&\coloneqq \sum_{x^T}(\alpha-Q_{X^T}(x^T))_+.
\end{align*}
Let $\calA\coloneqq\{x^T: Q_{X^T}(x^T)>\alpha\}$. We have
\begin{align}
    &F(P_{X^T})\geq \sum_{x^T:x^T\in\calA} \bigg(P_{X^T}(x^T)-\alpha\bigg)_+ \\
    &\geq \sum_{x^T:x^T\in\calA} \bigg(P_{X^T}(x^T)-\alpha\bigg)\\
    &=\sum_{x^T:x^T\in\calA} \bigg(P_{X^T}(x^T)-Q_{X^T}(x^T)+Q_{X^T}(x^T)-\alpha\bigg)\\
    &=E(Q_{X^T};\alpha)+\sum_{x^T:x^T\in\calA}(P_{X^T}(x^T)-Q_{X^T}(x^T)),
\end{align}
which leads to
\begin{align}
    &E(Q_{X^T};\alpha)-F(P_{X^T})\leq \sum_{x^T:x^T\in\calA}(Q_{X^T}(x^T)-P_{X^T}(x^T)) \nn\\
    &\leq \sum_{x^T:x^T\in\calA}(Q_{X^T}(x^T)-P_{X^T}(x^T))_+\\
    &\leq \sum_{x^T}(Q_{X^T}(x^T)-P_{X^T}(x^T))_+\leq d.
\end{align}
Thus, we have one lower bound for $F(P_{X^T})$, i.e.,
\begin{equation}
    F(P_{X^T})\geq  E(Q_{X^T};\alpha)-d. \label{eq: F(P) lb 1}
\end{equation}

On the other hand, we can derive another lower bound for $F(P_{X^T})$ using the fact that $F(P_{X^T})=1-\sum_{x^T}(P_{X^T}(x^T)\wedge \alpha)$. For each $x^T\in\calX^T$,
\begin{itemize}
    \item If $Q_{X^T}(x^T)\geq\alpha$, then $Q_{X^T}(x^T)\wedge \alpha=\alpha$ and $P_{X^T}(x^T)\wedge \alpha\leq \alpha=Q_{X^T}(x^T)\wedge \alpha$;
    \item If $Q_{X^T}(x^T)<\alpha$, then $Q_{X^T}(x^T)\wedge \alpha=Q_{X^T}(x^T)$ and $P_{X^T}(x^T)\wedge \alpha\leq \alpha=Q_{X^T}(x^T)+(\alpha-Q_{X^T}(x^T))=Q_{X^T}(x^T)\wedge \alpha+(\alpha-Q_{X^T}(x^T))_+$.
\end{itemize}
As a result, we have
\begin{align}
    P_{X^T}(x^T)\wedge \alpha &\leq Q_{X^T}(x^T)\wedge \alpha+(\alpha-Q_{X^T}(x^T))_+,~~\text{and}\\
     1-F(P_{X^T}) & =\sum_{x^T}(P_{X^T}(x^T)\wedge \alpha)\\
    &\leq \sum_{x^T}(Q_{X^T}(x^T)\wedge \alpha)+S(Q_{X^T};\alpha).
\end{align}
Therefore, $F(P_{X^T})$ has another lower bound as follows:
\begin{align}
    F(P_{X^T})&\geq 1-\sum_{x^T}(Q_{X^T}(x^T)\wedge \alpha)-S(Q_{X^T};\alpha)\\
    &=E(Q_{X^T};\alpha)-S(Q_{X^T};\alpha).  \label{eq: F(P) lb 2}
\end{align}
From \eqref{eq: F(P) lb 1} and \eqref{eq: F(P) lb 1} and the fact that $F(P_{X^T})\geq 0$, we have
\begin{equation}
    F(P_{X^T})\geq \bigg(E(Q_{X^T};\alpha)-\bigg(d \wedge S(Q_{X^T};\alpha)\bigg)\bigg)_+. \label{eq: F(P) final lb}
\end{equation}
There exists a distribution $P^*_{X^T}$ that can achieve this lower bound in \eqref{eq: F(P) final lb}. Intuitively, constructing $P^*_{X^T}$ is equivalent to the process of moving the mass of $Q_{X^T}$ above $\alpha$ to below $\alpha$, under some constraints. 

Define
\begin{align}
    \delta &\coloneqq \min\bigg\{d,\, S(Q_{X^T};\alpha),\, E(Q_{X^T};\alpha)\bigg\}, \\
    \calA&\coloneqq\{x^T: Q_{X^T}(x^T)>\alpha\}, \quad \calB\coloneqq\{x^T: Q_{X^T}(x^T)\leq\alpha\}.
\end{align}
Choose nonnegative transfers $\tau(x^T,y^T)\ge 0$ satisfying
\begin{align}
&\sum_{y^T\in \calB}\tau(x^T,y^T)\le Q_{X^T}(x^T)-\alpha,\\
&\sum_{x^T\in \calA}\tau(x^T,y^T)\le \alpha-Q_{X^T}(x^T),\\
&\sum_{x^T\in \calA}\sum_{y^T\in \calB}\tau(x^T,y^T)=\delta.
\end{align} 
Such $\tau$ can always be constructed using greedy algorithms. Then the optimal $P^*_{X^T}$ that achieves the lower bound in \eqref{eq: F(P) final lb} is given by
\begin{align}
    P^*_{X^T}(x^T)=\begin{cases}
        Q_{X^T}(x^T)-\sum_{y^T\in \calB}\;\tau(x^T,y^T), & x^T\in\calA,\\
        Q_{X^T}(x^T)+\sum_{y^T\in \calA}\;\tau(y^T,x^T), & x^T\in\calB.
    \end{cases} 
\end{align}
Note that $P^*_{X^T}$ is a valid probability pass function and satisfies $\rvD_\TV(P^*_{X^T},Q_{X^T})=\delta\leq d$.
By such construction, $P^*_{X^T}(x^T)\leq \alpha$ for all $x^T\in \calB$ and  $P^*_{X^T}(x^T)> \alpha$ for all $x^T\in \calA$. Then the induced $F(P^*_{X^T})$ is
\begin{align}
    F(P^*_{X^T})&=\sum_{x^T}(P^*_{X^T}(x^T)-\alpha)_+ = \sum_{x^T:x^T\in\calA}(P^*_{X^T}(x^T)-\alpha) \nn\\
    &=\sum_{x^T:x^T\in\calA}( Q_{X^T}(x^T)-\sum_{y^T\in \calB}\tau(x^T,y^T)-\alpha)\nn\\
    &=E(Q_{X^T};\alpha)-\delta.
\end{align}
If $\delta=E(Q_{X^T};\alpha)$, then $F(P^*_{X^T})=0$; otherwise, $F(P^*_{X^T})=E(Q_{X^T};\alpha)-\bigg(d \wedge S(Q_{X^T};\alpha)\bigg)$. Thus, $P^*_{X^T}$ achieves the lower bound in \eqref{eq: F(P) final lb}, proving optimality. In conclusion, under the distortion constraint defined with total variation distance, the lower bound for the detection error probability is given by the lower bound in \eqref{eq: F(P) final lb}.

\subsection{Validity of $P^*$ Constructed in Appendix \ref{App: pf of thm:cont-opt-construction}}\label{Supp: supp to pf of thm:cont-opt-construction}

We verify that $P^*$ is indeed a coupling of $P^*_{X^T}$ and $P_{\zeta^T}$.

\paragraph{Row sums.}
For each $x^T$,
\begin{align}
\int_{\calZ^T} P^*(x^T,\zeta^T) ~d \zeta^T
&=
P_{\zeta^T}(B(x^T)) + \frac{r(x^T)}{P_{\zeta^T}(B_0)} P_{\zeta^T}(B_0) \nn\\
&=
m(x^T)+r(x^T)
=
P^*(x^T).
\end{align}

\paragraph{Column sums.}
By disjointness of the sets $\{B(x^T)\}$, for any measurable set $A\subseteq \calZ^T$, 
\begin{align}
&\sum_{x^T} P^*(x^T,\zeta^T) \nn\\
&=\sum_{x^T}P_{\zeta^T}(A\cap B(x^T))+ \dfrac{\sum_{x^T}r(x^T)}{P_{\zeta^T}(B_0)}\,P_{\zeta^T}(A\cap B_0)\nn\\
&\overset{\text{(a)}}{=}\sum_{x^T}P_{\zeta^T}(A\cap B(x^T))+P_{\zeta^T}(A\cap B_0) \nn\\
&=P_{\zeta^T}(A), \nn
\end{align}
where (a) follows since 
\[
\sum_{x^T} r(x^T)
=
\sum_{x^T}\bigl(P^*(x^T)-m(x^T)\bigr)
=
1-\sum_{x^T} m(x^T)
=
P_{\zeta^T}(B_0).
\]

Hence $P^*$ has the desired marginals.

\subsection{Proof of Theorem \ref{thm:randomized-discrete-optimal-construction}}\label{App: pf of thm randomized-opt-cons}

\begin{restatedtheoremthree}[Restated]
Fix an LLM with generation distribution $Q_{X^T}$, $\alpha \in [0,1]$ and $\epsilon>0$. 
There exist a coupling $P_{X^T,\zeta^T}\in\Pi(P_{X^T}^*,P_{\zeta^T})$ and a randomized detector $\gamma$
that attains the lower bound in Lemma~\ref{Thm:min typeII error fixed} if and only if there exist subsets of $\calZ^T=\mathrm{supp}(P_{\zeta^T})$ satisfying the following conditions:
\begin{enumerate}
\item Pairwise disjoint sets $E_{x^T}\subseteq\calZ^T$, ${x^T}\in\calH$, satisfying
\begin{equation}
\alpha\le P_{\zeta^T}(E_{x^T})\le P_{X^T}^*(x^T),\qquad {x^T}\in\calH.
\label{eq:exclusive}
\end{equation}
\item With $\calR=\calZ^T\setminus\bigcup_{{x^T}\in\calH}E_{x^T}$, sets
$S_{x^T}\subseteq\calR$, ${x^T}\in\calL$, satisfying
\begin{equation}
P_{\zeta^T}(S_{x^T})\le\alpha,\qquad {x^T}\in\calL,
\label{eq:support-budget}
\end{equation}
and the capacity inequalities
\begin{equation}
\sum_{{x^T}\in A}P_{X^T}^*(x^T)\le P_{\zeta^T}\!\left(\bigcup_{{x^T}\in A}S_{x^T}\right),
\qquad A\subseteq\calL.
\label{eq:hall}
\end{equation}
\end{enumerate}
The sets $S_{x^T}$ need not be disjoint. When these conditions hold, a suitable
coupling can be constructed so that the detector
\begin{equation}
\gamma(1\mid x^T,\zeta^T)=
\begin{cases}
\displaystyle\frac{\alpha}{P_{\zeta^T}(E_{x^T})}\mathbbm{1}\{{\zeta^T}\in E_{x^T}\},&{x^T}\in\calH,\\[6pt]
\mathbbm{1}\{{\zeta^T}\in S_{x^T}\},&{x^T}\in\calL
\end{cases}
\label{eq:detector}
\end{equation}
attains the lower bound in Lemma \ref{Thm:min typeII error fixed}.
\end{restatedtheoremthree}

We first give the allocation fact used in the constructive direction, i.e., the sufficiency proof of Theorem~\ref{thm:randomized-discrete-optimal-construction}.

\begin{lemma}[Allocation with prescribed row masses and column capacities]
\label{lem:allocation}
Let $\calL$ be finite, let $S_{x^T}\subseteq\calR$, and let $P_{X^T}^*(x^T)\ge0$.
There is a nonnegative array $M(x^T,\zeta^T)$, ${x^T}\in\calL$, ${\zeta^T}\in\calR$, such that
\begin{equation}
\begin{gathered}
M(x^T,\zeta^T)=0\ \text{for }{\zeta^T}\notin S_{x^T},\\[3pt]
\sum_{{\zeta^T}\in\calR}M(x^T,\zeta^T)=P_{X^T}^*(x^T),\\[3pt]
\sum_{{x^T}\in\calL}M(x^T,\zeta^T)\le P_{\zeta^T}(\zeta^T)
\end{gathered}
\label{eq:allocation}
\end{equation}
if and only if \eqref{eq:hall} holds.
\end{lemma}
The proof of Lemma~\ref{lem:allocation} is deferred to the end. Lemma~\ref{lem:allocation} shows that there exists a nonnegative matrix $M$ that can couple the light masses of  $P_{X^T}$ supported on the sets $S_{x^T}$ and part of the masses from $P_{\zeta^T}$. Such matrix $M$ is used to construct the complete coupling between $P_{X^T}$ and $P_{\zeta^T}$ that achieves the optimality. 

Next, we present the proof of Theorem~\ref{thm:randomized-discrete-optimal-construction}
For any coupling and detector, define
\begin{align}
&a_{x^T}=\sum_{\zeta^T}P_{\zeta^T}(\zeta^T)\gamma(1\mid x^T,\zeta^T), \nn\\
&d_{x^T}=\sum_{\zeta^T}P_{X^T,\zeta^T}(x^T,\zeta^T)\gamma(1\mid x^T,\zeta^T). \nn
\end{align}
Since the null text distribution can be a point mass,
$\sup_{Q_{X^T}}\beta_0=\max_{x^T}a_{x^T}$. Also, $P_{X^T,\zeta^T}(x^T,\zeta^T)\le P_{\zeta^T}(\zeta^T)$ and $0\le \gamma(1\mid x^T,\zeta^T)\le1$
give
\begin{equation}
d_{x^T}\le P_{X^T}^*(x^T),\qquad d_{x^T}\le a_{x^T}\le\alpha.
\label{eq:bounds}
\end{equation}
Consequently, $\beta_1\ge1-\sum_{x^T}m(x^T)=\sum_{x^T}(P_{X^T}^*(x^T)-\alpha)_+$.
Because $\calX$ is finite, equality holds precisely when $d_{x^T}=m(x^T)$
for every ${x^T}$.

\noindent\textbf{Necessity ``$\Rightarrow$''.}
Suppose there exist a coupling \(P_{X^T,\zeta^T}\in\Pi(P_{X^T}^*,P_{\zeta^T})\) and a randomized detector \(\gamma\) that satisfy the worst-case Type-I error constraint and attain the Type-II error lower bound.
For ${x^T}\in\calH$, \eqref{eq:bounds} implies $d_{x^T}=a_{x^T}=\alpha$, so
\[
0=a_{x^T}-d_{x^T}=\sum_{\zeta^T}(P_{\zeta^T}(\zeta^T)-P_{X^T,\zeta^T}(x^T,\zeta^T))\gamma(1\mid x^T,\zeta^T).
\]
Every summand is nonnegative. Hence $\gamma(1\mid x^T,\zeta^T)>0$ implies $P_{X^T,\zeta^T}(x^T,\zeta^T)=P_{\zeta^T}(\zeta^T)$.
Define $E_{x^T}=\{{\zeta^T}\in\calZ:P_{X^T,\zeta^T}(x^T,\zeta^T)=P_{\zeta^T}(\zeta^T)\}=\{{\zeta^T}\in\calZ:P_{X^T|\zeta^T}(x^T|\zeta^T)=1\}$. These sets are pairwise
disjoint since $\sum_{x^T\in\mathcal V^T}P_{X^T|\zeta^T}(x^T|\zeta^T)=1$. 
Moreover,
\begin{align}
&\alpha=a_{x^T}=\sum_{{\zeta^T}\in E_{x^T}}P_{\zeta^T}(\zeta^T)\gamma(1\mid x^T,\zeta^T)\le P_{\zeta^T}(E_{x^T})\\
&=\sum_{{\zeta^T}\in E_{x^T}}P_{X^T,\zeta^T}(x^T,\zeta^T)\le P_{X^T}^*(x^T).
\end{align}
Thus \eqref{eq:exclusive} holds.

For ${x^T}\in\calL$, equality gives $d_{x^T}=P_{X^T}^*(x^T)$, and therefore
\[
0=P_{X^T}^*(x^T)-d_{x^T}=\sum_{\zeta^T}P_{X^T,\zeta^T}(x^T,\zeta^T)(1-\gamma(1\mid x^T,\zeta^T)).
\]
It follows that $\gamma(1\mid x^T,\zeta^T)=1$ whenever $P_{X^T,\zeta^T}(x^T,\zeta^T)>0$. Define
$S_{x^T}=\{{\zeta^T}:P_{X^T,\zeta^T}(x^T,\zeta^T)>0\}$. No such atom can belong to any high-row
exclusive set $E_{y^T}$, so $S_{x^T}\subseteq\calR$. Furthermore,
\[
P_{\zeta^T}(S_{x^T})\le\sum_{\zeta^T}P_{\zeta^T}(\zeta^T)\gamma(1\mid x^T,\zeta^T)=a_{x^T}\le\alpha.
\]
For any $A\subseteq\calL$, the column marginal constraint gives
\begin{align}
\sum_{{x^T}\in A}P_{X^T}^*(x^T)
&=\sum_{{\zeta^T}\in\bigcup_{{x^T}\in A}S_{x^T}}\sum_{{x^T}\in A}P_{X^T,\zeta^T}(x^T,\zeta^T)\\
&\le P_{\zeta^T}\!\left(\bigcup_{{x^T}\in A}S_{x^T}\right).
\end{align}
This proves \eqref{eq:support-budget} and \eqref{eq:hall}.

\noindent\textbf{Sufficiency ``$\Leftarrow$'': 1) construct the coupling.}
Suppose the stated sets exist. By Lemma~\ref{lem:allocation}, allocate
the low-row masses using an array $M$ supported on the sets $S_{x^T}$.
For ${\zeta^T}\in\calR$, define the remaining column capacities
\[
c_{\zeta^T}=P_{\zeta^T}(\zeta^T)-\sum_{{x^T}\in\calL}M(x^T,\zeta^T)\ge0.
\]
For ${x^T}\in\calH$, define the remaining row masses
\[
h_{x^T}=P_{X^T}^*(x^T)-P_{\zeta^T}(E_{x^T})\ge0.
\]
Their totals agree:
\begin{align}
&D_0:=\sum_{{x^T}\in\calH}h_{x^T}
\\
&=1-\sum_{{x^T}\in\calH}P_{\zeta^T}(E_{x^T})-\sum_{{x^T}\in\calL}P_{X^T}^*(x^T)
=\sum_{{\zeta^T}\in\calR}c_{\zeta^T}.
\label{eq:residual-total}\\
&=P_{\zeta^T}(\mathcal R)
	-\sum_{x^T\in\mathcal L}
	\sum_{\zeta^T\in\mathcal R}M(x^T,\zeta^T)\\
&=\left[1-\sum_{x^T\in\mathcal H}
	P_{\zeta^T}(E_{x^T})\right]
	-\sum_{x^T\in\mathcal L}P_{X^T}^*(x^T).
\end{align}
If $D_0>0$, set $N(x^T,\zeta^T)=h_{x^T}c_{\zeta^T}/D_0$ for ${x^T}\in\calH$, ${\zeta^T}\in\calR$.
If $D_0=0$, set $N(x^T,\zeta^T)=0$; all the corresponding row and column
residuals then vanish. Define
\begin{equation}
P_{X^T,\zeta^T}(x^T,\zeta^T)=
\begin{cases}
P_{\zeta^T}(\zeta^T),&{x^T}\in\calH,\ {\zeta^T}\in E_{x^T},\\
N(x^T,\zeta^T),&{x^T}\in\calH,\ {\zeta^T}\in\calR,\\
M(x^T,\zeta^T),&{x^T}\in\calL,\ {\zeta^T}\in\calR,\\
0,&\text{otherwise}.
\end{cases}
\label{eq:coupling}
\end{equation}
Every high row has mass $P_{\zeta^T}(E_{x^T})+h_{x^T}=P_{X^T}^*(x^T)$, and every low row has mass
$P_{X^T}^*(x^T)$. A column in $E_{x^T}$ is assigned entirely to its designated row;
a column in $\calR$ has mass $\sum_{{x^T}\in\calL}M(x^T,\zeta^T)+c_{\zeta^T}=P_{\zeta^T}(\zeta^T)$.
Thus $P_{X^T,\zeta^T}\in\Pi(P_{X^T}^*,P_{\zeta^T})$.

\noindent\textbf{Sufficiency ``$\Leftarrow$'': 2) verify the detector and errors.}
Recall the constructed acceptance probabilities:
\[
\gamma(1\mid x^T,\zeta^T)=
\begin{cases}
\displaystyle\frac{\alpha}{P_{\zeta^T}(E_{x^T})}
\mathbbm{1}\{\zeta^T\in E_{x^T}\},&x^T\in\calH,\\[6pt]
\mathbbm{1}\{\zeta^T\in S_{x^T}\},&x^T\in\calL.
\end{cases}
\]
For $x^T\in\calH$, the condition $P_{\zeta^T}(E_{x^T})\ge\alpha>0$
ensures that the denominator is positive and the ratio is at most one.
For $x^T\in\calL$, the rule is indicator-valued. Thus this is a valid
randomized detector, with
$\gamma(0\mid x^T,\zeta^T)=1-\gamma(1\mid x^T,\zeta^T)$.

\smallskip
\noindent\emph{Type-I error.}
Under the null hypothesis the joint distribution is
$Q_{X^T}\otimes P_{\zeta^T}$. For any null text distribution $Q_{X^T}$,
substitution of the detector gives
\begin{align}
&\beta_0(\gamma,Q_{X^T},P_{\zeta^T})\notag\\
&=\sum_{x^T}Q_{X^T}(x^T)
  \sum_{\zeta^T}P_{\zeta^T}(\zeta^T)
  \gamma(1\mid x^T,\zeta^T)\notag\\
&=\sum_{x^T\in\calH}Q_{X^T}(x^T)
  \sum_{\zeta^T}P_{\zeta^T}(\zeta^T)
  \frac{\alpha}{P_{\zeta^T}(E_{x^T})}
  \mathbbm{1}\{\zeta^T\in E_{x^T}\}\notag\\
&\quad+\sum_{x^T\in\calL}Q_{X^T}(x^T)
  \sum_{\zeta^T}P_{\zeta^T}(\zeta^T)
  \mathbbm{1}\{\zeta^T\in S_{x^T}\}\notag\\
&=\sum_{x^T\in\calH}Q_{X^T}(x^T)
  \frac{\alpha}{P_{\zeta^T}(E_{x^T})}
  P_{\zeta^T}(E_{x^T})\\
  &\qquad \quad +\sum_{x^T\in\calL}Q_{X^T}(x^T)P_{\zeta^T}(S_{x^T})\nn\\
&=\alpha\sum_{x^T\in\calH}Q_{X^T}(x^T)
  +\sum_{x^T\in\calL}Q_{X^T}(x^T)P_{\zeta^T}(S_{x^T})\nn\\
&\le\alpha\sum_{x^T\in\calH}Q_{X^T}(x^T)
  +\alpha\sum_{x^T\in\calL}Q_{X^T}(x^T)
  =\alpha. \label{eq:explicit-type-I}
\end{align}
The inequality uses $P_{\zeta^T}(S_{x^T})\le\alpha$ for all
$x^T\in\calL$, and the final equality uses the partition
$\calH\cup\calL=\calX$. Since this holds for every $Q_{X^T}$,
\[
\sup_{Q_{X^T}}\beta_0(\gamma,Q_{X^T},P_{\zeta^T})\le\alpha.
\]
In particular, the per-text false-positive probabilities are
$a_{x^T}=\alpha$ for $x^T\in\calH$ and
$a_{x^T}=P_{\zeta^T}(S_{x^T})\le\alpha$ for $x^T\in\calL$.

\smallskip
\noindent\emph{Type-II error.}
Under the alternative hypothesis, expectations are instead taken with
respect to the constructed coupling $P_{X^T,\zeta^T}$.
Substituting the same detector into the detection power gives
\begin{align}
&1-\beta_1(\gamma,P_{X^T,\zeta^T})\notag\\
&=\sum_{x^T,\zeta^T}P_{X^T,\zeta^T}(x^T,\zeta^T)
  \gamma(1\mid x^T,\zeta^T)\notag\\
&=\sum_{x^T\in\calH}\sum_{\zeta^T}
  P_{X^T,\zeta^T}(x^T,\zeta^T)
  \frac{\alpha}{P_{\zeta^T}(E_{x^T})}
  \mathbbm{1}\{\zeta^T\in E_{x^T}\}\notag\\
&\quad+\sum_{x^T\in\calL}\sum_{\zeta^T}
  P_{X^T,\zeta^T}(x^T,\zeta^T)
  \mathbbm{1}\{\zeta^T\in S_{x^T}\}\notag\\
&\overset{\text{(a)}}{=}\sum_{x^T\in\calH}
  \frac{\alpha}{P_{\zeta^T}(E_{x^T})}
  \sum_{\zeta^T\in E_{x^T}}P_{\zeta^T}(\zeta^T)
  +\sum_{x^T\in\calL}\sum_{\zeta^T\in S_{x^T}}M(x^T,\zeta^T)
  \notag\\
&\overset{\text{(b)}}{=}\sum_{x^T\in\calH}\alpha
  +\sum_{x^T\in\calL}P_{X^T}^*(x^T)
  =\sum_{x^T}\min\{P_{X^T}^*(x^T),\alpha\},
\label{eq:explicit-power}
\end{align}
where (a) uses two properties of \eqref{eq:coupling}: for $x^T\in\calH$ and $\zeta^T\in E_{x^T}$, $P_{X^T,\zeta^T}(x^T,\zeta^T)=P_{\zeta^T}(\zeta^T)$; for $x^T\in\calL$ and $\zeta^T\in S_{x^T}\subseteq\calR$, $P_{X^T,\zeta^T}(x^T,\zeta^T)=M(x^T,\zeta^T)$, and (b) also uses the support and row-sum properties of $M$:
\[
\sum_{\zeta^T\in S_{x^T}}M(x^T,\zeta^T)
=\sum_{\zeta^T\in\calR}M(x^T,\zeta^T)
=P_{X^T}^*(x^T),\qquad x^T\in\calL.
\]
Thus each high-probability text contributes $\alpha$ to the detection power, while each low-probability text contributes its full marginal mass $P_{X^T}^*(x^T)$. Finally,
\begin{align}
\beta_1(\gamma,P_{X^T,\zeta^T})
&=1-\sum_{x^T}\min\{P_{X^T}^*(x^T),\alpha\}\notag\\
&=\sum_{x^T}\left[P_{X^T}^*(x^T)
  -\min\{P_{X^T}^*(x^T),\alpha\}\right]\notag\\
&=\sum_{x^T}\bigl(P_{X^T}^*(x^T)-\alpha\bigr)_+.
\label{eq:explicit-type-II}
\end{align}
This attains the lower bound while satisfying the worst-case Type-I error constraint, completing the sufficiency argument.

\vspace{1em}
\noindent\textbf{Proof of Lemma~\ref{lem:allocation}:}
\noindent\textbf{Necessity.} Necessity follows by summing the row masses over any $A\subseteq\calL$:
all this mass must be placed in $\bigcup_{{x^T}\in A}S_{x^T}$, i.e., 
\begin{align}
\sum_{x^T\in A}P_{X^T}^*(x^T)
&=\sum_{x^T\in A}\sum_{\zeta^T\in\mathcal R}
M(x^T,\zeta^T)\\
&=\sum_{\zeta^T\in \cup_{x^T\in A}S_{x^T}}\sum_{x^T\in A}
M(x^T,\zeta^T)\\
&\le\sum_{\zeta^T\in \cup_{x^T\in A}S_{x^T}}P_{\zeta^T}(\zeta^T)\\
&=P_{\zeta^T}\left(\bigcup_{x^T\in A}S_{x^T}\right).
\end{align}

\noindent\textbf{Sufficiency.}
Group the atoms of $P_{\zeta^T}$ according to their admissible text sequences: for each $J\subseteq\mathcal L$, let $C_J=\{\zeta^T\in\mathcal R:\{x^T\in\mathcal L:\zeta^T\in S_{x^T}\}=J\}$ and $b_J=P_{\zeta^T}(C_J)$. These groups form a finite partition of $\mathcal R$, even when $\mathcal R$ is countably infinite. Set $D=\sum_{x^T\in\mathcal L}P_{X^T}^*(x^T)$ and construct a finite network with source-to-text capacities $P_{X^T}^*(x^T)$, text-to-group capacities $D+1$ whenever $x^T\in J$, and group-to-sink capacities $b_J$.

Any cut crossing a text-to-group edge has capacity greater than $D$. Otherwise, if $A$ denotes the texts on the source side, all groups adjacent to $A$ must also lie on that side, and the cut capacity is at least
\begin{align}
&D-\sum_{x^T\in A}P_{X^T}^*(x^T)+\sum_{J:J\cap A\ne\varnothing}b_J
\\
&=D-\sum_{x^T\in A}P_{X^T}^*(x^T)+P_{\zeta^T}\!\left(\bigcup_{x^T\in A}S_{x^T}\right)
\ge D.
\end{align}
The max-flow/min-cut theorem therefore guarantees a flow of value $D$. Let $f_{x^T,J}$ denote its text-to-group flows, setting $f_{x^T,J}=0$ when $x^T\notin J$. Since every source-to-text edge is saturated, these flows satisfy
\[
\sum_J f_{x^T,J}=P_{X^T}^*(x^T),\qquad \sum_{x^T\in\mathcal L}f_{x^T,J}\le b_J.
\]
For $\zeta^T\in C_J$ with $b_J>0$, define $M(x^T,\zeta^T)=f_{x^T,J}P_{\zeta^T}(\zeta^T)/b_J$, and set $M=0$ on groups with $b_J=0$. The support restriction follows from $f_{x^T,J}=0$ for $x^T\notin J$; summing within each group gives the prescribed row sums, while the group-capacity inequalities give the column bounds. Thus, such nonnegative $M$ exists.

\subsection{Proof of Theorem \ref{Thm: optimal detector}}\label{App: pf of Thm: optimal detector}
Given a detector $\gamma^*(X^T,\zeta^T)=\mathbbm{1}\{X^T=g(\zeta^T)\}$ and watermark-induced coupling $P_{X^T,\zeta^T}^*$ constructed in Theorem \ref{Thm: optimal detector}, the Type-I and Type-II errors are as follows.

\noindent\textbf{Worst-case Type-I error:} $\forall y^T\in\calV^T$,
\begin{align}
     & \bbE_{P_{\zeta^T}^*}[\mathbbm{1}\{y^T=g(\zeta^T)\}]=\sum_{\zeta^T}P_{\zeta^T}^*(\zeta^T)\mathbbm{1}\{y^T=g(\zeta^T)\}\\
    &=\sum_{\zeta^T}\sum_{x^T}P_{X^T,\zeta^T}^*(x^T,\zeta^T)\mathbbm{1}\{y^T=g(\zeta^T)\}\\
    &=P_{X^T}^*(y^T)\sum_{\zeta^T}   P_{\zeta^T|X^T}^*(\zeta^T|y^T)\mathbbm{1}\{y^T=g(\zeta^T)\} =P_{X^T}^*(y^T)\wedge \alpha \\
    &\leq \alpha,
\end{align}
and since any distribution $Q_{X^T}$ can be written as a linear combinations of $\delta_{y^T}$, we have $\max_{Q_{X^T}}\bbE_{Q_{X^T}P_{\zeta^T}^*}[\mathbbm{1}\{X^T=g(\zeta^T)\}]\leq \alpha$.

\noindent\textbf{Type-II error:}
\begin{align}
     &1-\bbE_{P^*_{X^T,\zeta^T}}[\mathbbm{1}\{X^T=g(\zeta^T)\}] \\
     &=1-\sum_{x^T}\sum_{\zeta^T}P_{X^T,\zeta^T}^*(x^T,\zeta^T) \mathbbm{1}\{x^T=g(\zeta^T)\}\\
    &=1-\sum_{x^T}P_{X^T}^*(x^T)\sum_{\zeta^T}   P_{\zeta^T|X^T}^*(\zeta^T|x^T)\mathbbm{1}\{x^T=g(\zeta^T)\} \\
    &=1-\sum_{x^T}\big(P_{X^T}^*(x^T)\wedge \alpha \big)\\
    &=\sum_{x^T:P_{X^T}^*(x^T)>\alpha}(P_{X^T}^*(x^T)- \alpha).
\end{align}
The optimality of $(\gamma^*,P_{X^T,\zeta^T}^*)$ is thus proved. We note that $\sum_{\zeta^T}P_{X^T,\zeta^T}^*\gamma(x^T,\zeta^T)=m(x^T)$ for all $x^T$, which is the if-and only-if-condition for attaining the optimal objective value of \eqref{Eq: opt-optimizable} and the universally minimum Type-II error simultaneously.

Compared to \cite[Theorem 3.2]{huang2023towards}, their proposed detector is equivalent to $\gamma(X^T,\zeta^T)= \mathbbm{1}\{X^T=\zeta^T\}$, where $\calZ^T=\calV^T\cup\{\tilde{\zeta}^T\}$ and $\tilde{\zeta}^T\notin \calV^T$, meaning that it belongs to $\Gamma^*$. 

\subsection{Proof of Lemma \ref{Lem: opt watermark when uniform}}\label{App: pf of Lem: opt watermark when uniform}

Consider $\gamma(X^T,\zeta^T)=\mathbbm{1}\{X^T=\zeta^T\}$ and $\calV^T \subseteq\calZ^T$, which is a model-agnostic detector. Let us first assume the distortion level $\epsilon=0$. The objective function of \eqref{Eq: opt fix gamma} becomes $ \beta_1(\gamma,P_{X^T,\zeta^T} ) =P_{X^T,\zeta^T}(X^T\ne\zeta^T)$, whose minimum is well-known as $\rvD_\TV(Q_{X^T},P_{\zeta^T})$, and the minimizer is given by
\begin{align}
    &P^*_{X^T,\zeta^T}(x^T,\zeta^T)\\
    &=\begin{cases}
        Q_{X^T}(x^T) \wedge P_{\zeta^T}(\zeta^T), &\text{if}\; x^T=\zeta^T;\\
        \frac{(Q_{X^T}(x^T)-P_{\zeta^T}(x^T))_+  (P_{\zeta^T}(\zeta^T)-Q_{X^T}(\zeta^T))_+}{\rvD_\TV(Q_{X^T},P_{\zeta^T})}, &\text{otherwise}.
    \end{cases}\label{Eq: TV minimizer}
\end{align}
This holds for any given pair of $(Q_{X^T},P_{\zeta^T})$. This watermarking scheme basically tries to force $X^T=\zeta^T$ as often as possible.
The worst-case Type-I error probability is given by:
\begin{align}
   &\sup_{Q_{X^T}}\bbE_{Q_{X^T}P_{\zeta^T}}[\mathbbm{1}\{X^T=\zeta^T\}]\\
   &=\sup_{Q_{X^T}}\sum_{x^T}Q_{X^T}(x^T)\frac{1}{|\calZ^T|}=\frac{1}{|\calZ^T|}\leq \alpha,
\end{align}
which satisfies the constraint in \eqref{Eq: opt fix gamma}. Therefore, when $\epsilon=0$, the construction of $P^*_{X^T,\zeta^T}$ is the optimal solution of \eqref{Eq: opt fix gamma}.

By allowing $\epsilon\geq 0$, the watermarked induced coupling $P_{X^T,\zeta^T}$ has marginals $P_{\zeta^T}$ and $P_{X^T}\in\{P_{X^T}:\rvD(P_{X^T}, Q_{X^T} )\leq \epsilon\}$. Therefore, the minimum Type-II error is the minimum total variation distance between the two marginals, i.e.,
\begin{align}
     &\beta_{1,\text{exp}}^*(\gamma,Q_{X^T},P_{\zeta^T},\alpha,\epsilon)=\min_{P_{X^T}:\rvD(P_{X^T}, Q_{X^T} )\leq \epsilon} \rvD_\TV(P_{X^T},P_{\zeta^T})\\
     &=\min_{\substack{P_{X^T}:\rvD(P_{X^T} , Q_{X^T} )\leq \epsilon
    }} \sum_{x^T}\bigg(P_{X^T}(x^T)-\frac{1}{|\calZ^T|} \bigg)_+.
\end{align}
By letting 
\begin{equation}
    P^\dagger_{X^T}\coloneqq\argmin_{\substack{P_{X^T}:\rvD(P_{X^T} , Q_{X^T} )\leq \epsilon
    }} \sum_{x^T}\bigg(P_{X^T}(x^T)-\frac{1}{|\calZ^T|} \bigg)_+,
\end{equation}
the corresponding optimal coupling becomes
\begin{align}
    &P^*_{X^T,\zeta^T}(x^T,\zeta^T)\\
    &=\begin{cases}
        P^\dagger_{X^T}(x^T) \wedge P_{\zeta^T}(\zeta^T), &\text{if}\; x^T=\zeta^T;\\
        \frac{(P^\dagger_{X^T}(x^T)-P_{\zeta^T}(x^T))_+  (P_{\zeta^T}(\zeta^T)-P^\dagger_{X^T}(\zeta^T))_+}{\rvD_\TV(P^\dagger_{X^T},P_{\zeta^T})}, &\text{otherwise}.
    \end{cases}\label{Eq: TV minimizer}
\end{align}

\end{document}